 \documentclass[]{jfm}
\usepackage{graphicx}
\usepackage{amsmath}%
\usepackage{subfig}
\usepackage{psfrag}
\usepackage[usenames, dvipsnames]{color}
\usepackage{float}
\usepackage{natbib}
\usepackage{pifont}

\ifCUPmtlplainloaded \else
  \checkfont{eurm10}
  \iffontfound
    \IfFileExists{upmath.sty}
      {\typeout{^^JFound AMS Euler Roman fonts on the system,
                   using the 'upmath' package.^^J}%
       \usepackage{upmath}}
      {\typeout{^^JFound AMS Euler Roman fonts on the system, but you
                   dont seem to have the}%
       \typeout{'upmath' package installed. JFM.cls can take advantage
                 of these fonts,^^Jif you use 'upmath' package.^^J}%
      }
  \else
  \fi
\fi

\ifCUPmtlplainloaded \else
  \checkfont{msam10}
  \iffontfound
    \IfFileExists{amssymb.sty}
      {\typeout{^^JFound AMS Symbol fonts on the system, using the
                'amssymb' package.^^J}%
       \usepackage{amssymb}%
       \let\le=\leqslant  
       \let\ge=\geqslant  
      }{}
  \fi
\fi

\ifCUPmtlplainloaded \else
  \IfFileExists{amsbsy.sty}
    {\typeout{^^JFound the 'amsbsy' package on the system, using it.^^J}%
     \usepackage{amsbsy}}
    {}
\fi

\newcommand\Rey{\mbox{\textit{Re}}}  % Reynolds number
\newcommand\Pran{\mbox{\textit{Pr}}} % Prandtl number, cf TeX's \Pr product
\newcommand\Nu{\mbox{\textit{Nu}}} % Nusselt number, cf TeX's \Pr product
\newcommand\Ri{\mbox{\textit{Ri}}} % Richardson number
\newcommand\Ra{\mbox{\textit{Ra}}} % Rayleigh number
\DeclareMathOperator\erf{erf}

\newcommand\solid[1][1.0cm]{\rule[0.5ex]{#1}{.4pt}}
\newcommand\ldash{\mbox{%
  \solid[2mm]\hspace{2mm}}}
\newcommand\dash{\mbox{%
  \solid[1mm]\hspace{1mm}}}
\newcommand\dashdot{\mbox{%
  \solid[1mm]$\cdot$}}
\newcommand\dashdotdot{\mbox{%
  \solid[1mm]$\cdot\cdot$}}

\newsavebox{\astrutbox}
\sbox{\astrutbox}{\rule[-5pt]{0pt}{20pt}}

\newcommand{\diff}{\mathrm{d}}
\newcommand\sqcf{\sqrt{\frac 2{C_f(\Rey_b)}}}
\newcommand\sqcfi{\sqrt{\frac {C_f(\Rey_b)}2}}

\title[]{Mixed convection in turbulent channels with unstable stratification}

\author[S. Pirozzoli, M. Bernardini, R. Verzicco, P. Orlandi]{SERGIO PIROZZOLI$^1$, MATTEO BERNARDINI$^1$, \\ ROBERTO VERZICCO$^{2,3}$ AND PAOLO ORLANDI$^1$} 

\affiliation{$^1$Dipartimento di Ingegneria Meccanica e Aerospaziale,
Sapienza Universit\`a di Roma \\ Via Eudossiana 18, 00184 Roma,
Italy \\ $^2$DII, Universit\`a di Roma Tor Vergata, Roma, Italy\\
$^3$PoF, University of Twente, Enschede, The Netherlands\\[\affilskip]}

\pubyear{1996}
\volume{538}
\pagerange{119--126}
\date{\today}

\begin{document}

\maketitle

\begin{abstract}
We study turbulent flows in planar channels with unstable thermal stratification,
using direct numerical simulations in a wide range of Reynolds and Rayleigh numbers and 
reaching flow conditions which are representative of 
asymptotic developed turbulence.
The combined effect of forced and free convection produces 
a peculiar pattern of quasi--streamwise rollers occupying the full 
channel thickness with aspect--ratio considerably higher than unity; it has been observed that they have an important redistributing effect
on temperature and momentum. The mean values and the variances of the flow variables
do not appear to follow Prandtl's scaling in the flow regime near free convection,
except for the temperature and vertical velocity fluctuations, which are more
affected by turbulent plumes. Nevertheless, we find that the Monin--Obukhov
theory still yields a useful representation of the main
flow features. In particular, the widely used Businger--Dyer relationships provide
a convenient way of accounting for the bulk effects of shear and buoyancy, 
although individual profiles may vary widely from the alleged trends.
Significant deviations are found in DNS with respect to the commonly used
parametrization of the mean velocity in the light-wind regime, which may have important
practical impact in models of atmospheric dynamics. 
Finally, for modelling purposes, we devise a set of empirical predictive formulas for
the heat flux and friction coefficients which can be used with about $10\%$ maximum
error in a wide range of flow parameters.

\end{abstract}

\section{Introduction}\label{sec:intro}

Mixed convection is the process whereby heat and momentum are transferred under the concurrent effect of 
shear and buoyancy, and it is at the heart of several physical phenomena of great 
practical importance, especially within the context of atmospheric dynamics.
Flow stratification may be either of stable type (i.e. the higher layers are hotter than the lower), 
or unstable type (i.e. lower layers are heated). In the former case, stratification suppresses 
the vertical motions thus mitigating friction and heat transfer. In contrast, 
unstable stratification 
promotes turbulent exchanges with obvious opposite effects. 
Although the two extreme cases of pure forced convection (classical boundary layers and channel flows)
and of free convection (Rayleigh-B\'enard flow) have been extensively studied 
analytically, experimentally and numerically, their combination appears to be much less understood.
The current engineering practice ~\citep{kays_80,incropera_11} for heat transfer prediction in the presence of mixed convection
mainly relies on correlations developed for either free or forced convection, and applied according to
the value of a global Richardson number, which grossly weights the effect of bulk
buoyancy with respect to shear. However, combination of the two effects can give rise to
flow patterns not recovered in either of the extreme cases.
For instance, it is known~\citep{avsec_37,hill_68,brown_80} than in the presence of unstable stratification
large streamwise-oriented rollers may form, which are regarded to be responsible
for the formation of aligned patterns of strato-cumulus clouds~\citep{mal_30,kuettner_59,kuettner_71}, for the 
peculiar striped patterns in desert sand dunes with typical spanwise spacing of a few kilometers~\citep{hanna_69},
and for the occurrence of long rows of unburned tree crowns in forest fires~\citep{haines_82}.
Wavy perturbations of the ordered pattern of convective rolls have also been frequently observed 
in the atmosphere~\citep{avsec_37,avsec_37_2}, and theoretically interpreted as the result of 
secondary instabilities~\citep{clever_91,clever_92}.

The current theoretical understanding of mixed convective flows heavily relies on the
framework set up by~\citet{obukhov_46,monin_54}. Mainly based on dimensional arguments, the
Monin-Obukhov theory relies on the existence of a universal length scale which incorporates 
the effects of friction and buoyancy, and defined as 
\begin{equation}
L = \frac{u_{\tau}^3}{Q \beta g}, \label{eq:MO}
\end{equation}
where $u_{\tau} = \sqrt{\tau_w/\rho}$ is the friction velocity (with $\tau_w$ and $\rho$ the time-- and surface--averages wall viscous stress and the fluid density, respectively), $Q$ is the total vertical heat flux, 
$\beta$ is the thermal expansion coefficient of the fluid, and $g$ is the gravity acceleration.
It should be noted that the definition of equation~\eqref{eq:MO} is consistent with that
by \citet{kader_90}, which specifically deals with the case of unstable stratification, 
although it slightly differs from the most common definition, which also includes a minus sign 
in front of the expression, and incorporates the Karman constant at the denominator. 
Given the definition of the Monin-Obukhov length scale, 
it is expected that for wall distances below $L$ convection dominates and 
a logarithmic (like) dependence of the flow variables on the wall distance
is observed.On the other hand, for distances from the wall beyond
$L$ buoyancy should dominate, with typical power--law scalings, to be discussed afterwards.
The theory is frequently used in the meteorological context to estimate stress and heat flux
from mean velocity and temperature gradients~\citep{stull_12}, and in large-eddy-simulation (LES) 
as a wall function for enforcement of numerical boundary conditions at off-wall locations~\citep{deardorff_72}.
The Monin-Obukhov theory has received substantial experimental confirmation
from atmospheric field measurements although with a large degree of scatter owing to 
inherent measurement uncertainties, and even certain basic features as the
asymptotic mean velocity and temperature scalings under light--wind conditions 
are the subject of current controversy~\citep{rao_04,rao_06}. 
Numerical support for Monin--Obukhov theory is to date rather scarce and it 
mainly comes from the large--eddy--simulation (LES) studies by \citet{khanna_97,khanna_98,johansson_01}, which 
raised issues regarding the proper scaling of horizontal velocity fluctuations. It is worthwhile noting that those LES studies were carried out
at realistic atmospheric conditions, hence possibly affected by approximate 
sub--grid--scale parametrization as well as by uncertainties incurred with
the use of Monin--Obukhov scaling to model the near--ground flow. 
To our knowledge, the Monin--Obukhov theory has never been systematically 
scrutinized through direct numerical simulation (DNS) and this is the main motivation of our study.

The channel flow is probably the most prototypical wall--bounded shear flow, 
and it has been extensively studied through DNS shed light 
on several important facets of wall turbulence structure. In particular, recent numerical studies have 
highlighted deviations from the alleged universal behavior of wall turbulence associated with 
high--Re effects~\citep{bernardini_14,lee_15}. DNS studies have also addressed the behaviour of 
passive scalars transported by the fluid phase, which serve to model dispersion of dilute contaminants 
as well as turbulent thermal transport under the assumption of small temperature differences. 
The latest studies~\citep{pirozzoli_16} have achieved a friction Reynolds 
number $\Rey_{\tau} \approx 4000$ (here $\Ra_{\tau} = u_{\tau} h / \nu$, 
where $h$ is the channel half--height) hence making it possible to establish 
the presence of a generalized logarithmic layer for the mean scalar profiles
although with a slightly different set of constants than those of the 
streamwise velocity. 
Classical predictive formulas for heat transfer based on the log law ~\citep{kader_81} are found to work
quite well, with a suitable choice of the constants.
%{\bf SERGIO un po' di tempo fa ho visto un presentazione di Moser su un canale a $Re_\tau =5000$; se fosse uscito qualcosa bisognerebbe citarlo.}

%METTERE LE REFERENZE DEL PERIODO DI SOTTO
At the opposite side of the flow parameter range the Rayleigh-B\'enard flow,
corresponding to free convection between isothermal walls, has been extensively
studied, both experimentally and numerically, mainly in cylindrical confined 
configurations~\citep{ahlers_09}.
Rayleigh numbers up to $10^{15}$ have been reached experimentally~\citep{niemela_00,ahlers_12} 
while fully resolved numerical simulations are behind at $\Ra = 2\times 10^{12}$ \citep{stevens_11}; around $\Ra=10^{14}$ the boundary 
layers adjacent to the heated plates become turbulent, and a transition to a 
state of ultimate convection is expected, marking a change from the classical 
$\Nu \sim \Ra^{1/3}$ behaviour to a steeper 
%SERGIO: SICURO DELLA FORMULA SOTTO?
$\sim \Ra^{(1/2)}/\ln(\Ra^{3/2})$ ($\simeq \Ra^{0.38}$ at $\Ra=10^{14}$).
Attempts are ongoing to push the Rayleigh number of the simulations beyond
$\Ra = 2\times 10^{12}$ since the experimental evidence in the ultimate regime is
not unanimous and reliable numerical simulations would help solving the 
controversy. 
Pure Rayleigh--B\'enard flow in planar geometry has to date received less attention,
probably because of the lack of matching experimental data, and uncertainties related to potential dependence on
the wall--parallel computational box size~\citep{hamman_15}, the most notable example dating back to \citet{kerr_96}.
Recent numerical simulations of Rayleigh--B\'enard flow in planar geometry in the presence 
of wall roughness and including the effect of finite thermal conduction at the walls have
been carried out by \citet{orlandi_15a}.

Numerical simulations and experiments of internal flows in the presence of buoyancy have been relatively infrequent so far.
Channel flows with stable temperature stratification have received some attention in recent years, 
with important contributions from DNS delivered by~\citet{armenio_02,garcia_11}, leading to
the conclusion that flow relaminarization in the channel core may be achieved depending on the value of the bulk Richardson number. 
However, as pointed out by \citet{garcia_11},
this effect sensitively depends on the size of the computational box, and extremely large 
domains are required to achieve box--independent results, even for the lowest order statistics.
Channels with unstable stratification were considered in the experiments of \citet{mizushina_82,fukui_85},
which showed that for strongly unstable stratification the flow pattern is radically different than in the 
neutrally buoyant case, owing to the onset of quasi-longitudinal rollers which make up for large part of
momentum and heat transfer. In one of the few quantitative measurements~\citep{fukui_91}
it was found that the typical size of each roller is about $1.3$ times the channel height,
although the aspect ratio of the duct was about $7$, hence effects are likely to be significant.
The formation of longitudinal rollers in turbulent flows was first confirmed in the DNS of \citet{domaradzki_88},
which however was dealing with Couette flow between sliding plates. 
\citet{iida_97} carried out numerical simulations of plane channel flows with unstable 
stratification at low bulk Richardson numbers ($\Ri_b \le 0.3$) and low Reynolds number
($\Rey_{\tau} = 150$), finding steady increase of
the Nusselt number with $\Ri_b$, whereas (quite interestingly) the friction coefficient was found to slightly
decrease up to $\Ri_b \approx 0.05$. 
\citet{sid_15} extended the envelope of the flow parameters to $\Rey_{\tau} \lesssim 400$, $\Ri_b \lesssim 1$,
confirming the non--monotonic trend of friction, and observing a typical blunting of the 
velocity and temperature profiles as the effect of buoyancy becomes significant.
Using a DNS database at $\Rey_{\tau} \lesssim 200$, $\Ra \lesssim 10^7$, \citet{scagliarini_15} 
developed a phenomenological model resulting in a modified logarithmic law for the mean velocity 
which incorporates the effects of shear and buoyancy.
Non--Oberbeck--Boussinesq effects associated with temperature dependent fluid properties were also
studied by \citet{zonta_14}.
Notably, all the above numerical studies were carried out at rather low Reynolds and/or Rayleigh number,
hence they are not necessarily representative of fully turbulent flow conditions of practical 
relevance, and they span a limited range of Richardson numbers, typically close to the
case of pure forced convection. Further, with the exception of the work of \citet{zonta_14},
simulations have been mainly carried out in narrow channels, 
which may prevents natural self--organization of large--scale coherent structures, as shown
in recent studies on box size sensitivity~\citep{hamman_15}.
In fact, linear stability analysis~\citep{gage_68} predicts the onset on exponentially growing disturbances in
the form of longitudinal rolls in the presence of unstable stratification, even at very low Rayleigh number.
The most unstable disturbances predicted by linear stability analysis have a typical spanwise wavelength of about $4 h$, which is comparable with the typical wavelength
of the rollers recovered in fully turbulent simulations. Hence, $L_z = 4 h$ may be regarded as a minimal spanwise box size to 
achieve healthy turbulence in channel flow simulations of mixed convection.

It is the main purpose of this study to establish a high--fidelity database 
for unstably buoyant channel flows 
which encompasses a wide range of Richardson numbers, at high enough values of Reynolds and 
Rayleigh number to be representative of fully developed turbulence. The numerical database is presented in \S\ref{sec:numerics}, 
the flow organization is discussed in \S\ref{sec:flow}, and the main flow statistics in \S\ref{sec:stats}.
The implications of the present findings in the light of Monin--Obukhov theory are discussed is \S\ref{sec:MO}, 
and considerations on heat transfer and friction predictive formulas are given in \S\ref{sec:heat}.

\section{The numerical database}\label{sec:numerics}

\begin{table}
 \centering
 \begin{tabular*}{1.\textwidth}{@{\extracolsep{\fill}}lcccccccccc}
  \hline
   Flow case          & $\Rey_b$ & $\Ra$   & $\Ri_b$    &  $h/L$        & $\Rey_{\tau}$ & $\Nu$ & $C_f$ & $N_x$ & $N_y$ & $N_z$ \\
  \hline
   RUN\_Ra9\_Re0        &    0     & $10^9$  & $\infty$ &  $\infty$     & 0       & 63.172  &   NA         & 6144  & 768   & 3072  \\  
   RUN\_Ra9\_Re4.5      & 31623    & $10^9$  & $1$      &  4.44264      & 946.41  & 60.255  &  7.16E-3     & 6144  & 768   & 3072  \\  
   RUN\_Ra8\_Re0        &    0     & $10^8$  & $\infty$ &  $\infty$     & 0       & 30.644  &   NA         & 2560  & 512   & 1280  \\  
   RUN\_Ra8\_Re3        &  1000    & $10^8$  & $100$    &  214.136      & 96.166  & 30.470  &  7.39E-2     & 2560  & 512   & 1280  \\  
   RUN\_Ra8\_Re3.5      &  3162    & $10^8$  & $10$     &  30.0932      & 179.12  & 27.672  &  2.56E-2     & 2560  & 512   & 1280  \\  
   RUN\_Ra8\_Re4        & 10000    & $10^8$  & $1 $     &  3.67709      & 351.01  & 25.443  &  9.85E-3     & 2560  & 512   & 1280  \\  
   RUN\_Ra8\_Re4.5      & 31623    & $10^8$  & $0.1$    &  0.44134      & 864.24  & 45.584  &  5.97E-3     & 2560  & 512   & 1280  \\  
   RUN\_Ra7\_Re0        &    0     & $10^7$  & $\infty$ &  $\infty$     & 0       & 15.799  &   NA         & 1024  & 256   &  512  \\  
   RUN\_Ra7\_Re2.5      &  316.2   & $10^7$  & $100   $ &  167.716      & 38.690  & 15.541  &  1.20E-1     & 1024  & 256   &  512  \\  
   RUN\_Ra7\_Re3        &  1000    & $10^7$  & $10    $ &  24.4576      & 70.992  & 14.000  &  4.03E-2     & 1024  & 256   &  512  \\  
   RUN\_Ra7\_Re3.5      &  3162    & $10^7$  & $1     $ &  3.01888      & 134.98  & 11.880  &  1.46E-2     & 1024  & 256   &  512  \\  
   RUN\_Ra7\_Re3.5\_LA  &  3162    & $10^7$  & $1     $ &  2.99729      & 136.12  & 12.094  &  1.48E-2     & 2048  & 256   & 1024  \\  
   RUN\_Ra7\_Re3.5\_SM  &  3162    & $10^7$  & $1     $ &  2.85494      & 136.60  & 11.642  &  1.49E-2     &  512  & 256   &  256  \\  
   RUN\_Ra7\_Re3.5\_NA  &  3162    & $10^7$  & $1     $ &  2.50569      & 142.59  & 11.622  &  1.62E-2     &  256  & 256   &  128  \\  
   RUN\_Ra7\_Re4        & 10000    & $10^7$  & $0.1   $ &  0.37257      & 307.01  & 17.250  &  7.54E-3     & 1024  & 256   &  512  \\  
   RUN\_Ra7\_Re4.5      & 31623    & $10^7$  & $0.01  $ &  0.04243      & 823.19  & 37.871  &  5.42E-3     & 2560  & 512   & 1280  \\  
   RUN\_Ra6\_Re0        &    0     & $10^6$  & $\infty$ &  $\infty$     & 0       & 8.2884  &   NA         &  512  & 192   &  256  \\  
   RUN\_Ra6\_Re2        &   100    & $10^6$  & $100   $ &  114.745      & 16.436  & 8.1528  &  2.16E-1     &  512  & 192   &  256  \\  
   RUN\_Ra6\_Re2.5      &  316.2   & $10^6$  & $10    $ &  16.0776      & 30.527  & 7.3180  &  7.45E-2     &  512  & 192   &  256  \\  
   RUN\_Ra6\_Re3        &  1000    & $10^6$  & $1     $ &  1.94472      & 58.894  & 6.3560  &  2.77E-2     &  512  & 192   &  256  \\  
   RUN\_Ra6\_Re3.5      &  3162    & $10^6$  & $0.1   $ &  0.29783      & 112.47  & 6.7801  &  1.02E-2     &  512  & 192   &  256  \\  
   RUN\_Ra6\_Re4        & 10000    & $10^6$  & $0.01  $ &  0.02906      & 298.92  & 12.419  &  7.15E-3     & 1024  & 256   &  512  \\  
   RUN\_Ra6\_Re4.5      & 31623    & $10^6$  & $0.001 $ &  0.00349      & 817.63  & 30.508  &  5.35E-3     & 2560  & 512   & 1280  \\  
   RUN\_Ra5\_Re3.5      &  3162    & $10^5$  & $0.01  $ &  0.02262      & 108.48  & 4.6190  &  9.41E-3     &  512  & 192   &  256  \\  
   RUN\_Ra5\_Re4        & 10000    & $10^5$  & $0.001 $ &  0.00284      & 297.86  & 12.013  &  7.10E-3     & 1024  & 256   &  512  \\  
   RUN\_Ra4\_Re3        &  1000    & $10^4$  & $0.01  $ &  0.01508      & 45.731  & 2.3073  &  1.67E-2     &  512  & 192   &  256  \\  
   RUN\_Ra4\_Re3.5      &  3162    & $10^4$  & $0.001 $ &  0.00225      & 107.22  & 4.4345  &  9.19E-3     &  512  & 192   &  256  \\  
   RUN\_Ra0\_Re3.5      &  3162    &   0     & 0        &  0            & 106.78  & 4.4836  &  9.12E-3     &  512  & 192   &  256  \\  
   RUN\_Ra0\_Re4        & 10000    &   0     & 0        &  0            & 297.78  & 12.009  &  7.09E-3     & 1024  & 256   &  512  \\  
   RUN\_Ra0\_Re4.5      & 31623    &   0     & 0        &  0            & 815.60  & 29.757  &  5.32E-3     & 2560  & 512   & 1280  \\  
  \hline
 \end{tabular*}
 \caption{List of parameters for buoyant turbulent channel DNS. $\Rey_b = 2 h u_b / \nu$ is the bulk Reynolds number, $\Rey_{\tau} = h u_{\tau} / \nu$ is the friction Reynolds number, $\Ri_b=2 \beta g \Delta \theta h / u_b^2$ is the bulk Richardson number, $\Ra=\beta g \Delta \theta (2 h)^3 / (\alpha \nu)$ is the Rayleigh number, $\Nu= 2 h Q / (\alpha \Delta \theta)$ is the Nusselt number, $C_f=2 \tau_w / (\rho u_b^2)$ is the skin friction coefficient. $N_x$, $N_y$, $N_z$ are the number of grid points in the streamwise, wall-normal, and spanwise directions. An error stretching function $y (\eta) = \erf \left [ a \left (\eta - 0.5 \right ) \right] / \erf \left(0.5 \, a \right)$, $a = 3.2$, $\eta = [-1,1]$ has been used to cluster grid points in the wall-normal direction. All simulations are carried out in a $L_x \times L_z = 16 h \times 8 h$ box, except for those labeled as LA ($32 h \times 16 h$), SM ($8 h \times 4 h$), NA ($4 h \times 2 h$).}
 \label{tab:params}
\end{table}

\begin{table}
 \centering
%\begin{tabular*}{1.\textwidth}{@{\extracolsep{\fill}}cccc}
 \begin{tabular*}{0.5\textwidth}{@{\extracolsep{\fill}}lclc}
  \hline
   $\Rey_b$ & Line type  & $\Ra$ & Symbol type \\
  \hline
  0          & \color{Gray}{$\cdot\cdot\cdot\cdot\cdot\cdot\cdot$} & 0        & $\square$ \\
   $10^{2.5}$ & \color{Orange}{\dash\dash\dash\dash\solid[1mm]}       & $10^5$   & $\Delta$  \\
   $10^{3  }$ & \color{Green}{\dashdot\dashdot\dashdot\dashdot\dashdot} & $10^6$  & $\nabla$  \\
   $10^{3.5}$ & \color{Red}{\dashdotdot\dashdotdot\dashdotdot\solid[1mm]}   & $10^7$  & {\Large $\triangle$}  \\
   $10^{4  }$ & \color{Blue}{\ldash\ldash\solid[2mm]} & $10^8$  & {\Large $\diamond$}  \\
   $10^{4.5}$ & \color{Black}{\solid}  & $10^9$  & {\huge $\circ$}  \\ 
  \hline
 \end{tabular*}
 \caption{Nomenclature of lines (indicating the value of $\Ra$) and symbols (indicating the value of $\Rey_b$.}
 \label{tab:nomen}
\end{table}

\begin{figure}
 \centering
  \psfrag{x}[t][][1.0]{$\Ra$}
  \psfrag{y}[b][][1.0]{$\Rey_b$}
  \includegraphics[width=6.0cm,clip]{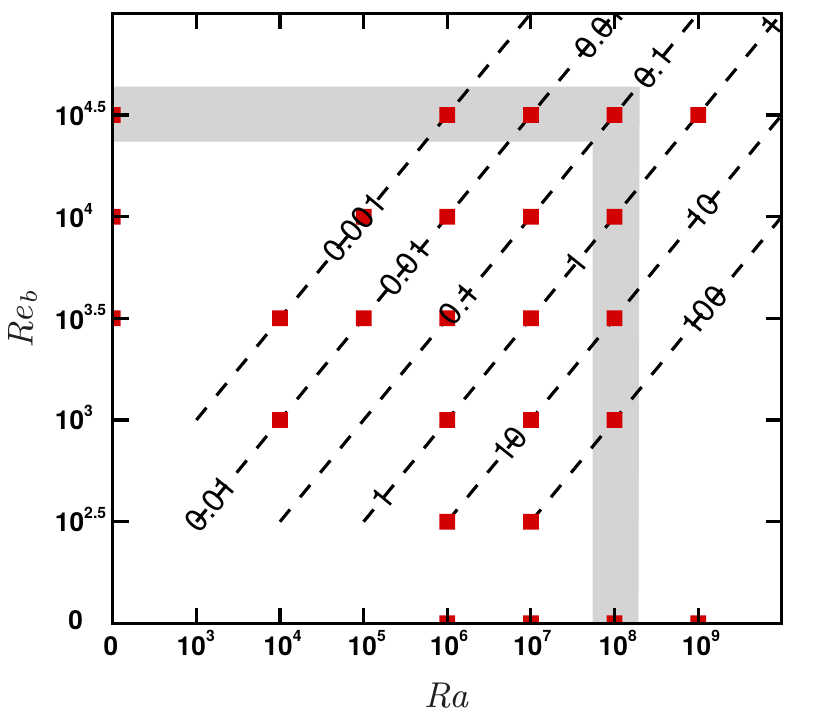}
  \caption{Overview of flow cases in the $\Ra$-$\Rey_b$ plane. Each solid square corresponds to a DNS. The dashed diagonal lines have constant $\Ri_b$. The shaded area highlights the subset of DNS which have been mainly used for the statistical analysis. Note that the bi-logarithmic diagram is intentionally inconsistent for the purpose of including the extreme cases of free convection ($\Rey_b=0$) and forced convection ($\Ra=0$).}
  \label{fig:mesh}
\end{figure}

The Navier-Stokes equations for an incompressible buoyant fluid under the Boussinesq approximation are numerically solved
\begin{eqnarray}
\frac{\partial u_j}{\partial x_j} = 0, \quad \frac{\partial u_i}{\partial t} + \frac{\partial u_i u_j}{\partial x_j} &=&
-\frac{\partial p}{\partial x_i} + \beta g \theta \delta_{i2} + \Pi \delta_{i1} + \nu \frac{\partial^2 u_i}{\partial x_j \partial x_j}, \label{eq:momentum} \\
\frac{\partial \theta}{\partial t} + \frac{\partial \theta u_j}{\partial x_j} &=&
\alpha \frac{\partial^2 \theta}{\partial x_j \partial x_j}, \label{eq:scalar}
\end{eqnarray}
where $u_i$ are the Cartesian velocity components  
($i=1,2,3$ corresponding, respectively to the streamwise, wall--normal, and spanwise directions), 
$\theta$ is the temperature perturbation with respect to a reference state of hydrostatic equilibrium,
%FORSE BISOGNEREBBE DIRE QUALE E' QUESTO MEAN STATE
$\beta$ is the thermal expansion coefficient, $g$ is the gravity acceleration, 
$\Pi = u_{\tau}^2/h$ is the forcing acceleration used to maintain a constant mass flow rate, and
$\nu$ and $\alpha$ are the  kinematic viscosity and temperature diffusivity, respectively.
For that purpose, an existing channel flow solver with passive scalar transport~\citep{pirozzoli_16} has been modified; the code
is capable to discretely preserve the total kinetic energy and the scalar variance
in the limit of vanishing diffusivities and time integration error.

The present flow is controlled by three global parameters, namely the bulk Reynolds number $\Rey_b = 2 h u_b / \nu$,
where $u_b$ is the channel bulk velocity, the Rayleigh number, $\Ra = (8 h^3 \beta g \Delta \theta)/(\alpha \nu)$,
and the Prandtl number, $\Pran = \nu/\alpha$, where $\Delta \theta$ is the temperature difference
across the two walls. The relative importance of gravity as compared 
to convection is quantified by the bulk Richardson number, $\Ri_b = 2 h \beta g \Delta \theta / u_b^2 = \Ra /(\Rey_b^2 \Pran)$. 
All the simulations have been carried out at unit Prandtl number, covering the range of 
Reynolds and Rayleigh numbers $0 \le \Rey_b \le 10^{4.5} = 31623$, $0 \le \Ra \le 10^9$.
The various DNS are labeled according to the following convention: RUN\_Rax\_Rey denotes a 
run carried out at $\Ra = 10^x$, $\Rey_b = 10^y$, $\Ra=0$ corresponding to pure Poiseuille flow, and $\Rey_b=0$
corresponding to pure Rayleigh-B\'enard flow. Accordingly, the Richardson number may attain the 
extreme values $0$ and $\infty$, and finite values in the range $10^{-3}-10^2$, in multiples of 10.
An overview of the computed flow cases is presented in figure~\ref{fig:mesh} in the $\Ra$-$\Rey_b$ plane.
The extreme cases of Rayleigh-B\'enard and Pouiseuille flow are shown on the horizontal
and vertical axes, respectively. It is important to note that in the chosen doubly-logarithmic representation 
the iso-lines of the bulk Richardson number are diagonal lines, and the flow conditions are selected
in such a way that several flow cases are encountered along the iso-$\Ri_b$ lines to help isolate
the effects of the parameters into play.
Table \ref{tab:params} provides a list of the bulk flow parameters obtained for all the simulations herein carried out.
The grid spacings have been carefully selected in such a way that the resolution requirements put forth
by \citet{shishkina_10} are satisfied in the limit case of pure buoyant flow, and 
the spacings in wall units are $\Delta x^+ = \Delta z^+ \lesssim 4.5$ in pure Poiseuille flow~\citep{bernardini_14}.
The adequacy of the mesh resolution has been checked a-posteriori for all the simulations, and the 
grid size in each coordinate direction is nowhere larger that three local Kolmogorov units.

\begin{figure}
 \centering
  \includegraphics[width=12.0cm,clip]{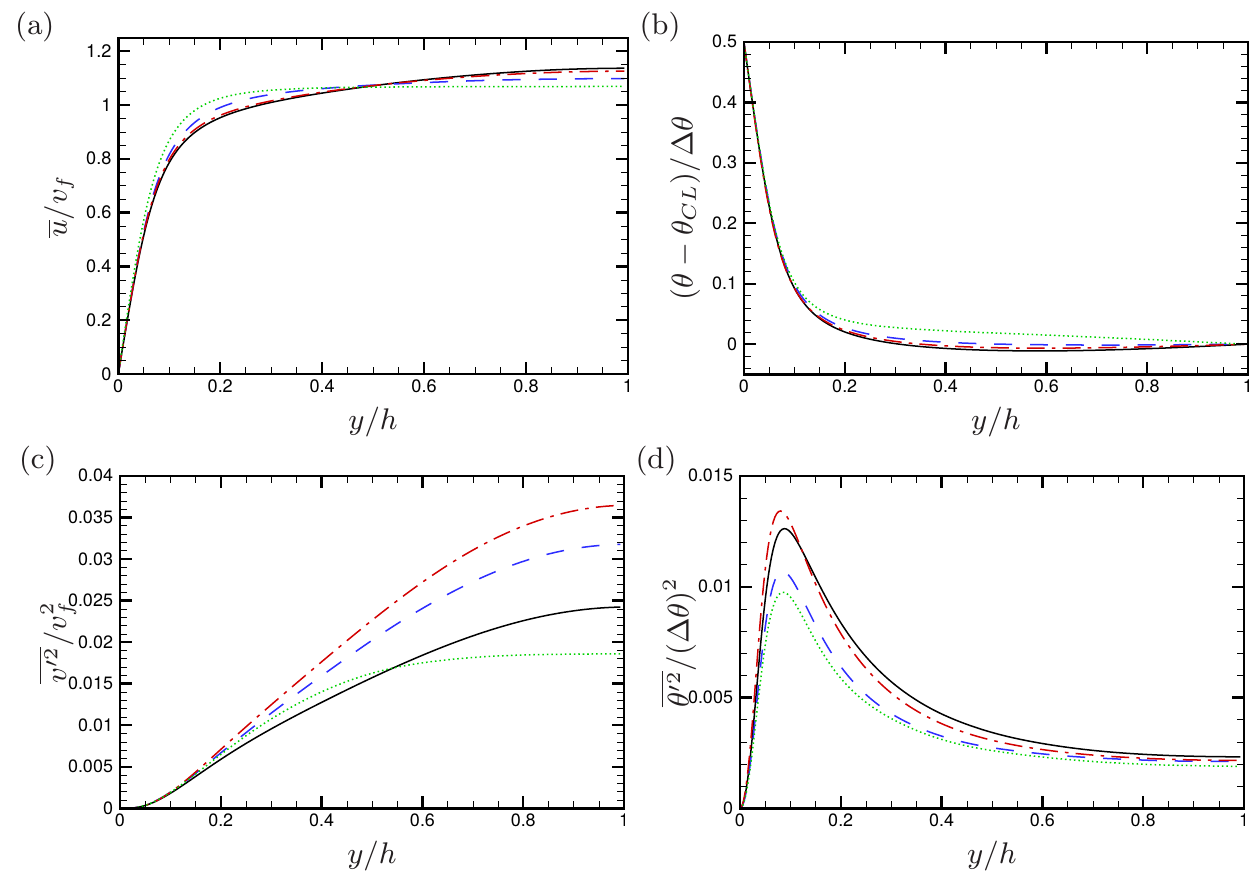}
  \caption{Sensitivity to computational box size for flow case  RUN\_Ra7\_Re3.5 ($\Ri_b=1$): (a) mean velocity; (b) mean temperature; 
  (c) variance of wall-normal velocity; (d) variance of temperature. 
  $v_f = (2 \beta g h \Delta \theta)^{1/2}$ is the reference free-fall velocity.
  Statistics are obtained in boxes 
  with $L_x \times L_z$ = $4 h \times 2 h$ (dotted lines); $8 h \times 4 h$ (dashed lines); 
  $16 h \times 8 h$ (solid lines); $32 h \times 16 h$ (dot-dashed lines).}
 \label{fig:boxsens}
\end{figure}

Preliminary simulations have been carried out for flow case RUN\_Ra7\_Re3.5 (having $\Ri_b = 1$) to
establish the effect of computational box size on the turbulence statistics. 
The bulk flow parameters for these DNS, listed in table~\ref{tab:params} suggest little
effect of box size, except for the smallest box (having $L_x=4h$, $L_z=2h$), which 
shows symptoms of severe numerical confinement. 
Differences are clearer in the statistics of velocity and temperature, as shown in figure~\ref{fig:boxsens}.
The figure suggests that the computational domain has effects even on the mean flow properties, 
and especially the velocity variable that tends to have a flatter spatial distribution in narrow domains. 
Near insensitivity of mean velocity and temperature is observed starting at $L_x \times L_z = 16 h \times 8 h$,
although velocity and temperature variances are still varying, even in non-monotonic fashion. 
Hence, given the need to keep the computational expense within reasonable bounds, and 
given the restrictions on the computational box size in Poiseuille flow~\citep{bernardini_14},
all simulations have been performed in a $16 h \times 8 h$ box.
 
For ease of later reference, the style of lines and symbols used to denote the various flow cases 
is explained in table~\ref{tab:nomen}.
To avoid possible confusion it is important to note that, consistent with (most of) the wall turbulence community,
the streamwise, wall-normal and spanwise coordinates are here labeled as $x$, $y$, $z$, respectively,
and the corresponding velocity components as $u$, $v$, $w$. In contrast, in the geophysical community
$z$ and $w$ are typically reserved for the vertical, wall normal,  direction.

\section{Flow organization} \label{sec:flow}

\begin{figure}
 \centering
% (a)
% \includegraphics[width=3.0cm,angle=270,clip]{DATA/RB/plotxzu_CL.eps}
  (a)
  \includegraphics[width=2.6cm,angle=270,clip]{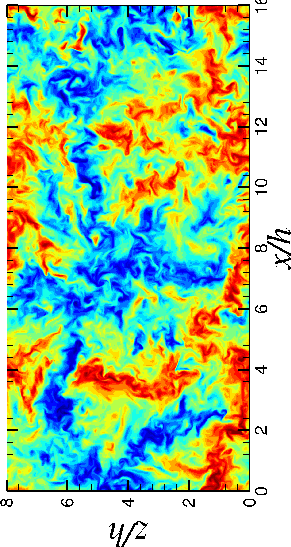}
  (b)
  \includegraphics[width=2.6cm,angle=270,clip]{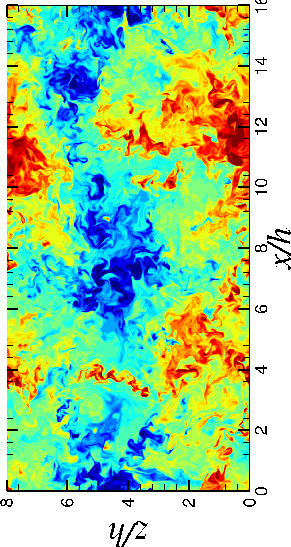}
  \\ \vskip0.25em
% (d)
% \includegraphics[width=3.0cm,angle=270,clip]{DATA/Ri100_Ra8/plotxzu_CL.png}
  (c)
  \includegraphics[width=2.6cm,angle=270,clip]{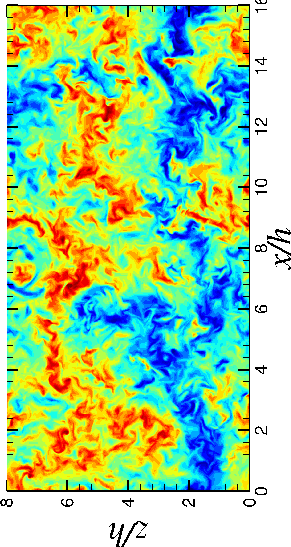}
  (d)
  \includegraphics[width=2.6cm,angle=270,clip]{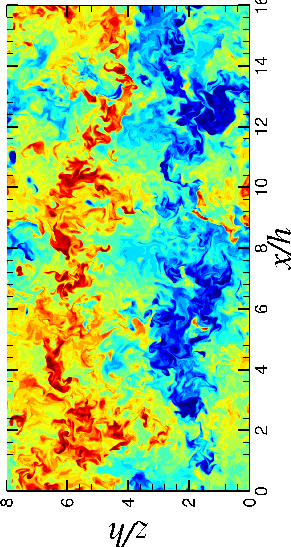}
  \\ \vskip0.25em
% (g)
% \includegraphics[width=3.0cm,angle=270,clip]{DATA/Ri10/plotxzu_CL.png}
% (h)
% \includegraphics[width=3.0cm,angle=270,clip]{DATA/Ri10/plotxzv_CL.png}
% (i)
% \includegraphics[width=3.0cm,angle=270,clip]{DATA/Ri10/plotxzT_CL.png}
% \\ \vskip1.em
% (g)
% \includegraphics[width=3.0cm,angle=270,clip]{DATA/Ri1/plotxzu_CL.png}
  (e)
  \includegraphics[width=2.6cm,angle=270,clip]{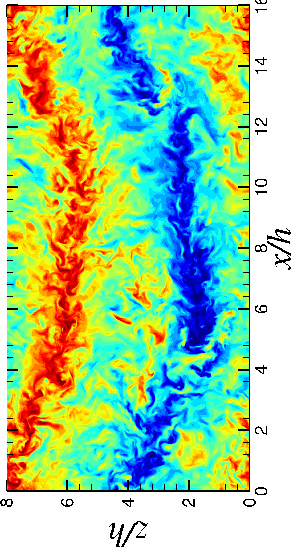}
  (f)
  \includegraphics[width=2.6cm,angle=270,clip]{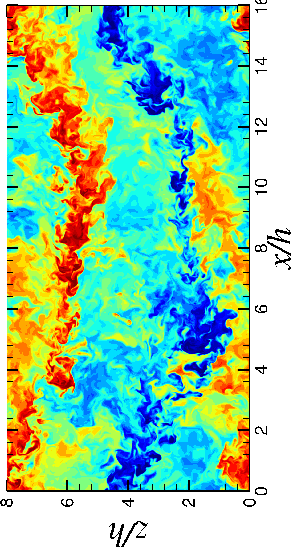}
  \\ \vskip0.25em
% (a)
% \includegraphics[width=3.0cm,angle=270,clip]{DATA/Ri0.1/plotxzu_CL.png}
% (b)
% \includegraphics[width=3.0cm,angle=270,clip]{DATA/Ri0.1/plotxzv_CL.png}
% (c)
% \includegraphics[width=3.0cm,angle=270,clip]{DATA/Ri0.1/plotxzT_CL.png}
% \\ \vskip1.em
% (d)
% \includegraphics[width=3.0cm,angle=270,clip]{DATA/Ri0.01_Ra7/plotxzu_CL.png}
  (g)
  \includegraphics[width=2.6cm,angle=270,clip]{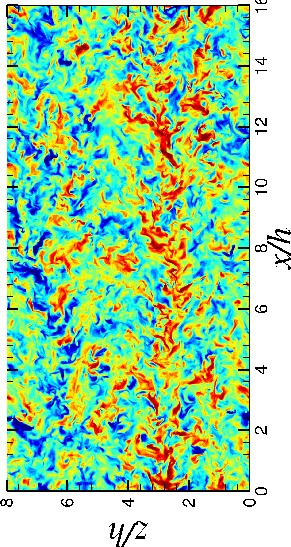}
  (h)
  \includegraphics[width=2.6cm,angle=270,clip]{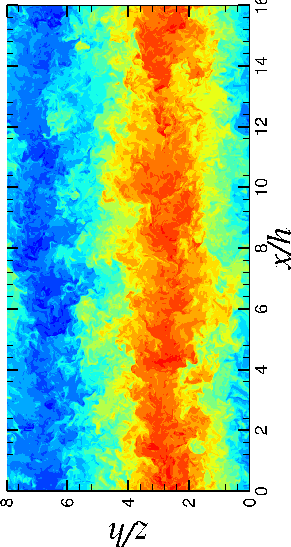}
  \\ \vskip0.25em
% (g)
% \includegraphics[width=3.0cm,angle=270,clip]{DATA/Ri0.001_Ra6/plotxzu_CL.png}
  (i)
  \includegraphics[width=2.6cm,angle=270,clip]{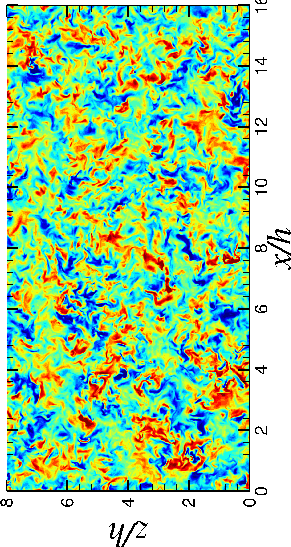}
  (j)
  \includegraphics[width=2.6cm,angle=270,clip]{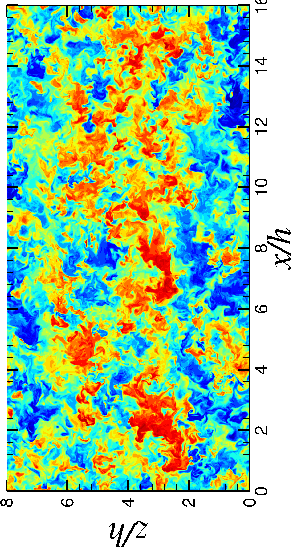}
  \\ \vskip0.25em
% (j)
% \includegraphics[width=3.0cm,angle=270,clip]{DATA/POISEUILLE_Reb4.5/plotxzu_CL.png}
  (k)
  \includegraphics[width=2.6cm,angle=270,clip]{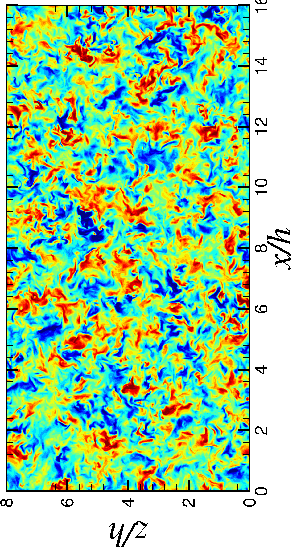}
  (l)
  \includegraphics[width=2.6cm,angle=270,clip]{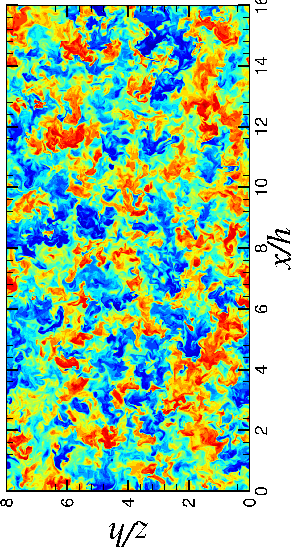}
  \\ \vskip0.25em
% \label{fig:Ra_xz_CL}
  \caption{Effect of Reynolds and Rayleigh number variation: instantaneous visualizations of $v'$ (left column) and $\theta'$ (right column) at channel center plane ($y=h$) for flow cases
 RUN\_Ra8\_Re0 ($\Ri_b=\infty$, first row), RUN\_Ra8\_Re3 ($\Ri_b=100$, second row), RUN\_Ra8\_Re4 ($\Ri_b=1$, third row), RUN\_Ra7\_Re4.5 ($\Ri_b=0.01$, fourth row), RUN\_Ra6\_Re4.5 ($\Ri_b=0.001$, fifth row), RUN\_Ra0\_Re4.5 ($\Ri_b=0$, sixth row). 24 contour levels are shown for each variable in the range $\pm 3$ standard deviations from the mean value (negative values in blue and positive in red).}
  \label{fig:Re_Ra_xz_CL}
\end{figure}

\begin{figure}
 \centering
  (a)
  \includegraphics[width=2.6cm,angle=270,clip]{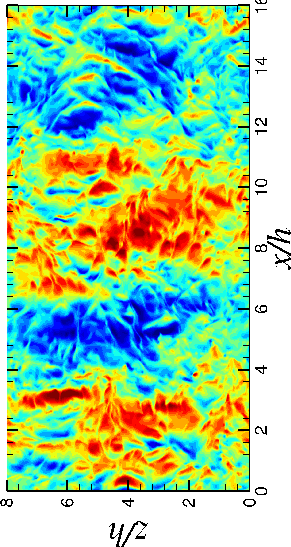}
  (b)
  \includegraphics[width=2.6cm,angle=270,clip]{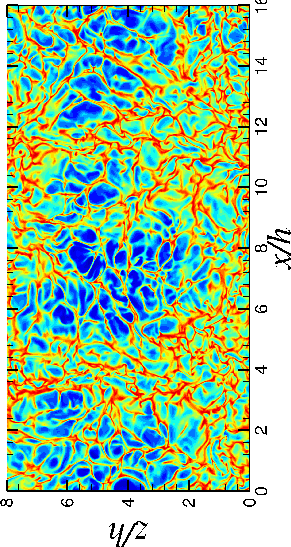}
  \\ \vskip0.25em
  (c)
  \includegraphics[width=2.6cm,angle=270,clip]{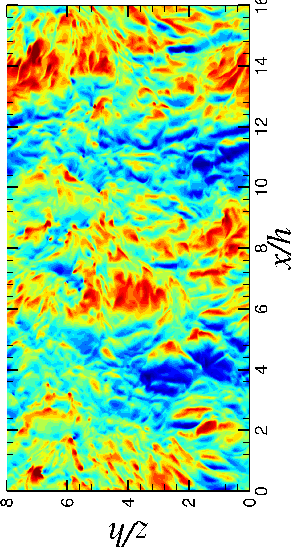}
  (d)
  \includegraphics[width=2.6cm,angle=270,clip]{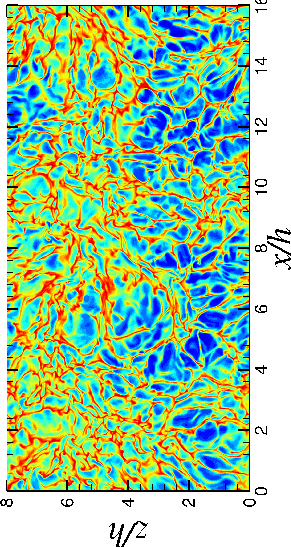}
  \\ \vskip0.25em
  (e)
  \includegraphics[width=2.6cm,angle=270,clip]{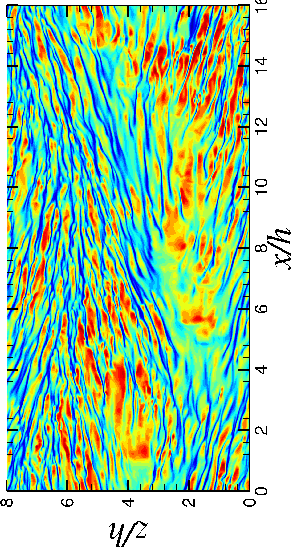}
  (f)
  \includegraphics[width=2.6cm,angle=270,clip]{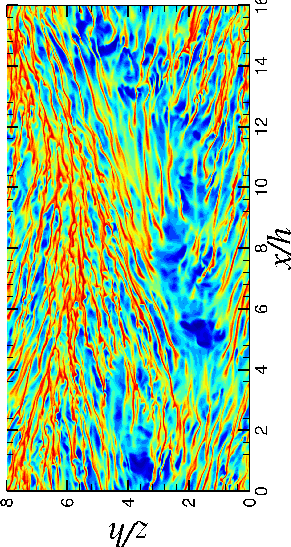}
  \\ \vskip0.25em
  (g)
  \includegraphics[width=2.6cm,angle=270,clip]{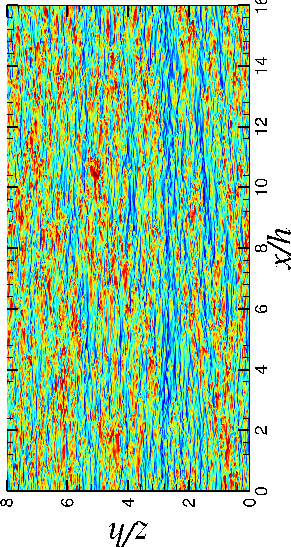}
  (h)
  \includegraphics[width=2.6cm,angle=270,clip]{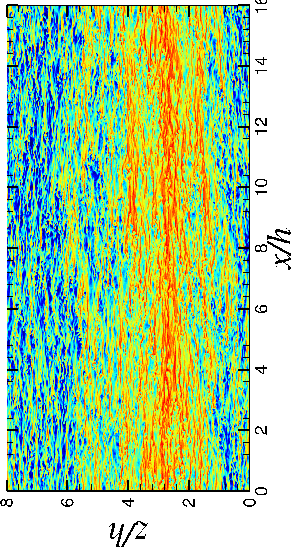}
  \\ \vskip0.25em
  (i)
  \includegraphics[width=2.6cm,angle=270,clip]{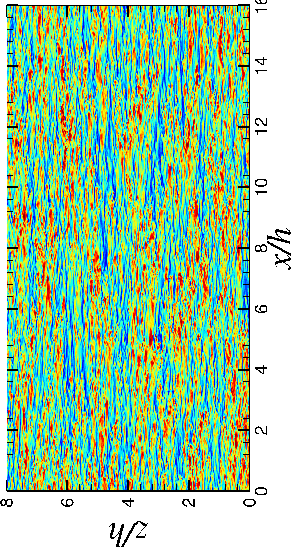}
  (j)
  \includegraphics[width=2.6cm,angle=270,clip]{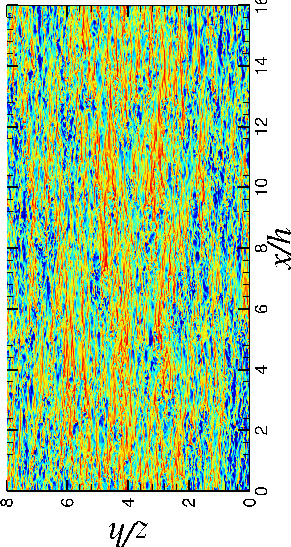}
  \\ \vskip0.25em
  (k)
  \includegraphics[width=2.6cm,angle=270,clip]{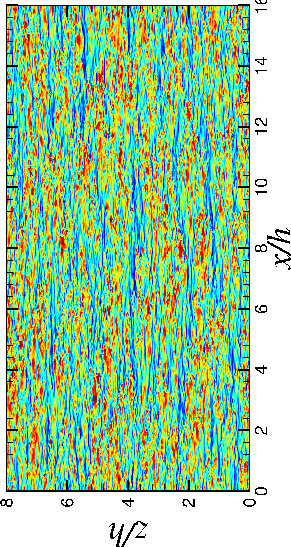}
  (l)
  \includegraphics[width=2.6cm,angle=270,clip]{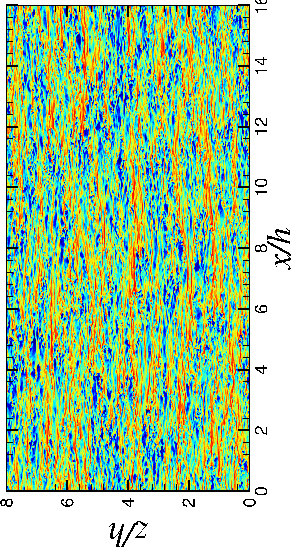}
  \\ \vskip0.25em
  \caption{Effect of Reynolds and Rayleigh number variation: instantaneous visualizations of $u'$ (left column) and $\theta'$ (right column) at near-wall station ($y=y_P$) for flow cases
 RUN\_Ra8\_Re0 ($\Ri_b=\infty$, first row), RUN\_Ra8\_Re3 ($\Ri_b=100$, second row), RUN\_Ra8\_Re4 ($\Ri_b=1$, third row), RUN\_Ra7\_Re4.5 ($\Ri_b=0.01$, fourth row), RUN\_Ra6\_Re4.5 ($\Ri_b=0.001$, fifth row), RUN\_Ra0\_Re4.5 ($\Ri_b=0$, sixth row). 24 contour levels are shown for each variable in the range $\pm 3$ standard deviations from the mean value (negative values in blue and positive in red).}
  \label{fig:Re_Ra_xz_NW}
\end{figure}

\begin{figure}
 \centering
  \centerline{
  (a)
  \includegraphics[width=1.8cm,angle=270,clip]{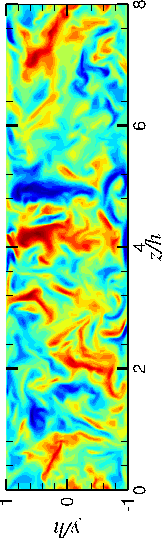}
  (b)
  \includegraphics[width=1.8cm,angle=270,clip]{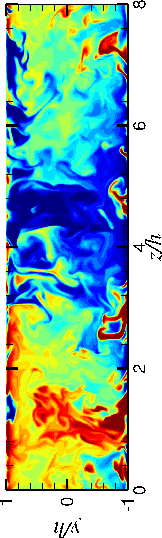}
  }
  \vskip1.em
  \centerline{
  (c)
  \includegraphics[width=1.8cm,angle=270,clip]{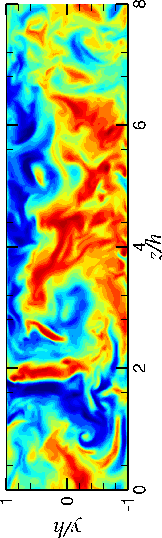}
  (d)
  \includegraphics[width=1.8cm,angle=270,clip]{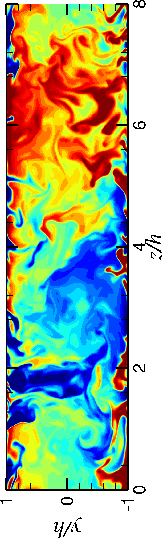}
  } 
  \vskip1.em
  \centerline{
  (e)
  \includegraphics[width=1.8cm,angle=270,clip]{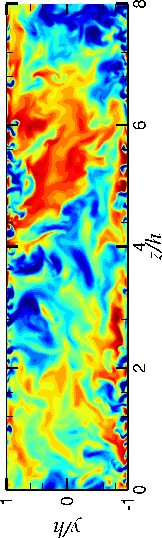}
  (f)
  \includegraphics[width=1.8cm,angle=270,clip]{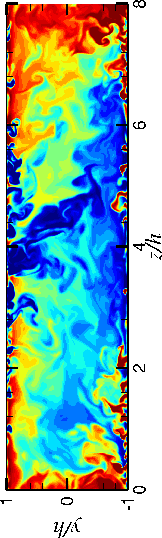}
  } 
  \vskip1.em
  \centerline{
  (g)
  \includegraphics[width=1.8cm,angle=270,clip]{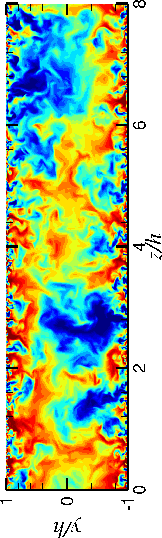}
  (h)
  \includegraphics[width=1.8cm,angle=270,clip]{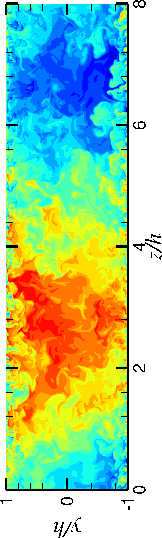}
  } 
  \vskip1.em
  \centerline{
  (i)
  \includegraphics[width=1.8cm,angle=270,clip]{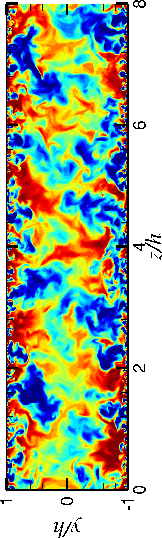}
  (j)
  \includegraphics[width=1.8cm,angle=270,clip]{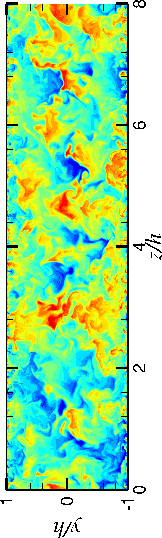}
  } 
  \vskip1.em
  \centerline{
  (k)
  \includegraphics[width=1.8cm,angle=270,clip]{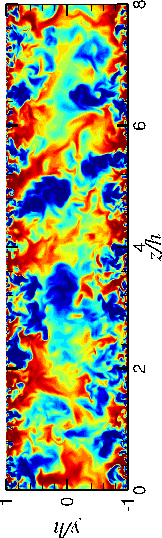}
  (l)
  \includegraphics[width=1.8cm,angle=270,clip]{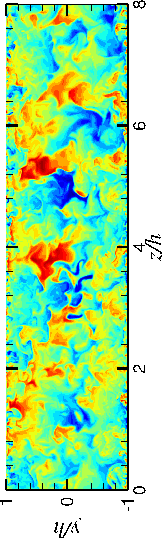}
  } 
  \vskip1.em
  \caption{Effect of Reynolds and Rayleigh number variation: instantaneous visualizations of $v'$ (left column) and $\theta'$ (right column) in cross-stream plane for flow cases
 RUN\_Ra8\_Re0 ($\Ri_b=\infty$, first row), RUN\_Ra8\_Re3 ($\Ri_b=100$, second row), RUN\_Ra8\_Re4 ($\Ri_b=1$, third row), RUN\_Ra7\_Re4.5 ($\Ri_b=0.01$, fourth row), RUN\_Ra6\_Re4.5 ($\Ri_b=0.001$, fifth row), RUN\_Ra0\_Re4.5 ($\Ri_b=0$, sixth row). 24 contour levels are shown for each variable in the range $\pm 3$ standard deviations from the mean value (negative values in blue and positive in red).}
  \label{fig:Re_Ra_yz}
\end{figure}

The flow structure is scrutinized in this section by analyzing instantaneous snapshots of the flow variables and 
their corresponding spectral densities. To get insights into the combined effects of Reynolds and Rayleigh numbers, 
we show velocity and temperature fluctuations
(figures~\ref{fig:Re_Ra_xz_CL}--\ref{fig:Re_Ra_yz}) 
in wall--parallel and cross--stream planes for several cases along the outer edge of the parameter--space matrix,
marked with a shaded area in figure~\ref{fig:mesh}, for decreasing $\Ri_b$.
Specifically, DNS results are presented at constant $\Ra=10^8$ for increasing $\Rey_b$, up to $\Ri_b=1$,
and then at constant $\Rey_b=10^{4.5}$ for decreasing $\Ra$, down to the limit case of Poiseuille flow.
Two representative wall distances have been selected for the analysis;
the channel centerline ($y=h$), where $v'$ and $\theta'$ probe the large--scale flow organization, 
and the position of peak production of temperature fluctuations ($P_{\theta}=-\overline{v'\theta'} \diff \overline{\theta} / \diff y$), 
hereafter indicated as $y=y_P$, where we show $u'$ and $\theta'$.
In the limit of free convection (panels (a)--(b)), persistent large--scale flow organization is observed at the channel centerline, consisting of a network of rollers
which transport hot fluid from the bottom to the top (and vice versa)
through upward-- and downward--traveling plumes, as is evident from figure~\ref{fig:Re_Ra_yz}(a),(b).
The flow visualizations suggest that the rollers have axes preferably pointing in the $x$ and in the $z$ 
direction, which is a likely consequence of the rectangular geometry of the computational domain.
Indeed, the footprint of rollers with axes aligned in the $z$ direction is apparent in the near--wall 
distribution of $u$ (see figure~\ref{fig:Re_Ra_xz_NW}a).
As soon as the mean flow--rate becomes non--zero (panels (c)--(d), corresponding to $\Ri_b=100$), the flow attains a
more definite spatial orientation, the rollers now mostly pointing in the streamwise direction.
It is noteworthy that only two pairs of counter--rotating rollers are captured within the selected computational box.
At intermediate Richardson numbers a strong meandering behavior of the rollers is observed, which is most evident at $\Ri_b=1$ (panels (e)--(f)), 
and which is the likely result of wavy instability of the counter--rotating rollers. 
This kind of instability is frequently observed in the meteorological context~\citep{avsec_37, benard_38},
and in experiments~\citep{pabiou_05},
and it was theoretically explained in the context of laminar flow by \citet{clever_91}.
The association between vertical velocity and temperature fluctuations 
at the channel centreline, which well reflects the importance of vertical motions
in the redistribution of the temperature field. 
Waviness of the rollers seems to be suppressed at further low Richardson number ($\Ri_b=0.01$, see panels g,h),
at which the rollers are very nearly straight, and maximum organization is observed in cross-stream planes.
Loss of coherence of the rollers and loss of correlation between $v'$ and $\theta'$ is observed starting from 
$\Ri=0.001$ (panels (i)--(j)),
which marks the passage to a spotty organization in the channel core typical of Poiseuille flow.
In cross--stream planes (see figure~\ref{fig:Re_Ra_yz}), the change of regime from $\Ri_b=0.01$ to $\Ri_b=0.001$
is marked by the disappearance of rollers spanning the whole channel, although large eddies 
are still observed in the form of wall--attached ejections confined to each channel half~\citep{bernardini_14}.

As a result of the change in the bulk flow organization, the near--wall turbulence is also modified. 
In the case of free convection (figure~\ref{fig:Re_Ra_xz_NW}(a)--(b)) the typical temperature pattern observed in 
Rayleigh--B\'enard flow is
recovered, with a distinctive network of near--wall line plumes protruding from the boundary layer 
into the bulk flow~\citep{kerr_96}. A similar organization is found up to $\Ri_b = 1$ (panels (e)--(f)),
although a clear modulating influence from the overlaying rollers is found
as the most intense near--wall plumes tend to be embedded within large--scale updrafts which
are the ascending branch of core rollers (compare with figure~\ref{fig:Re_Ra_xz_CL}(e)--(f)).
In this regime the streamwise velocity does not have a definite small--scale organization, nor it
is clearly associated with the temperature field.
The scenario changes at $\Ri_b=0.01 $ (figure~\ref{fig:Re_Ra_xz_NW}(g)--(h)), with
momentum streaks first appearing near the wall, which have strongly negative correlation
with temperature fluctuations. At this $\Ri_b$, the streaks appear to be strongly modulated
by the action of the core rollers. However, as this action ceases ($\Ri_b \le 0.001$),
near--wall turbulence attains the typical organization of canonical wall--bounded flows.

%\begin{figure}
% \centering
%  (a)
%  \includegraphics[width=2.0cm,angle=270,clip]{DATA/Ri1_Ra9/plotxzu_NW.png}
%  (b)
%  \includegraphics[width=2.0cm,angle=270,clip]{DATA/Ri1_Ra9/plotxzv_NW.png}
%  (c)
%  \includegraphics[width=2.0cm,angle=270,clip]{DATA/Ri1_Ra9/plotxzT_NW.png}
%  \\ \vskip1.em
%  (d)
%  \includegraphics[width=2.0cm,angle=270,clip]{DATA/Ri1/plotxzu_NW.png}
%  (e)
%  \includegraphics[width=2.0cm,angle=270,clip]{DATA/Ri1/plotxzv_NW.png}
%  (f)
%  \includegraphics[width=2.0cm,angle=270,clip]{DATA/Ri1/plotxzT_NW.png}
%  \\ \vskip1.em
%  (g)
%  \includegraphics[width=2.0cm,angle=270,clip]{DATA/Ri1_Ra7/plotxzu_NW.png}
%  (h)
%  \includegraphics[width=2.0cm,angle=270,clip]{DATA/Ri1_Ra7/plotxzv_NW.png}
%  (i)
%  \includegraphics[width=2.0cm,angle=270,clip]{DATA/Ri1_Ra7/plotxzT_NW.png}
%  \\ \vskip1.em
%  \caption{Simulations at $\Ri_b=1$: instantaneous visualizations of $u'$ (left column), $v'$ (middle column), $\theta'$ (right column) at near-wall station ($y=y_P$) for flow case RUN\_Ra9\_Re4.5 (first row), RUN\_Ra8\_Re4 (second row), RUN\_Ra7\_Re3.5 (third row). 24 contour levels are shown for each variable in the range $\pm 3$ standard deviations from the mean value (negative values in blue and positive in red).}
%  \label{fig:Ri_xz_NW}
%\end{figure}

\begin{figure}
 \centering
  \centerline{
  (a)
  \includegraphics[width=2.0cm,angle=270,clip]{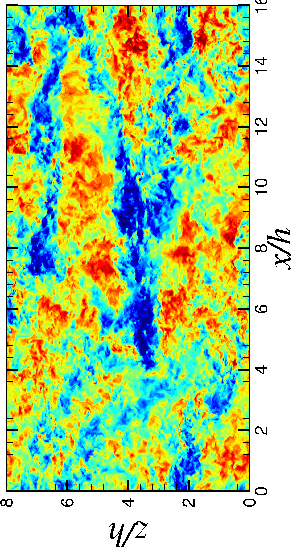}
  (b)
  \includegraphics[width=2.0cm,angle=270,clip]{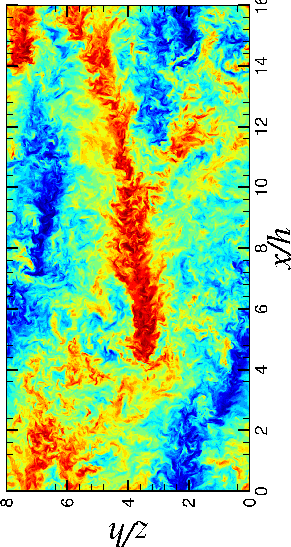}
  (c)
  \includegraphics[width=2.0cm,angle=270,clip]{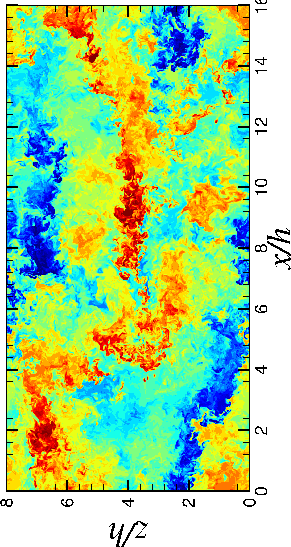}
  }
  \vskip1.em
  \centerline{
  (d)
  \includegraphics[width=2.0cm,angle=270,clip]{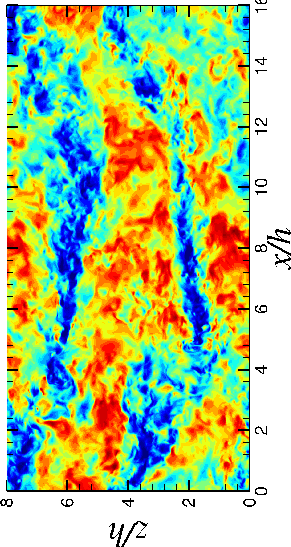}
  (e)
  \includegraphics[width=2.0cm,angle=270,clip]{DATA/Ri1/plotxzv_CL.png}
  (f)
  \includegraphics[width=2.0cm,angle=270,clip]{DATA/Ri1/plotxzT_CL.png}
  }
  \vskip1.em
  \centerline{
  (g)
  \includegraphics[width=2.0cm,angle=270,clip]{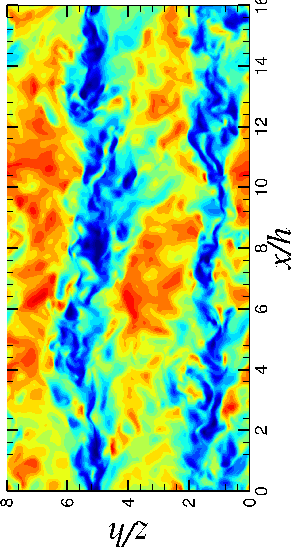}
  (h)
  \includegraphics[width=2.0cm,angle=270,clip]{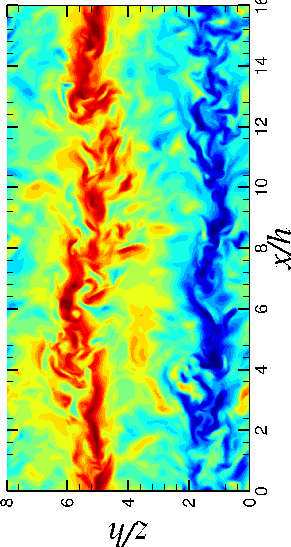}
  (i)
  \includegraphics[width=2.0cm,angle=270,clip]{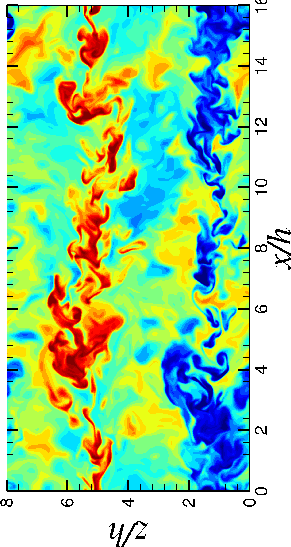}
  }
  \vskip1.em
  \caption{Numerical simulations at $\Ri_b=1$: instantaneous visualizations of $u'$ (left column), $v'$ (middle column), $\theta'$ (right column) at channel centerline for flow case RUN\_Ra9\_Re4.5 (first row), RUN\_Ra8\_Re4 (second row), RUN\_Ra7\_Re3.5 (third row). 24 contour levels are shown for each variable in the range $\pm 3$ standard deviations from the mean value (negative values in blue and positive in red).}
  \label{fig:Ri_xz_CL}
\end{figure}

%\begin{figure}
% \centering
%  (a)
%  \includegraphics[width=1.1cm,angle=270,clip]{DATA/Ri1_Ra9/plotyzu.png}
%  (b)
%  \includegraphics[width=1.1cm,angle=270,clip]{DATA/Ri1_Ra9/plotyzv.png}
%  (c)
%  \includegraphics[width=1.1cm,angle=270,clip]{DATA/Ri1_Ra9/plotyzT.png}
%  \\ \vskip1.em
%  (d)
%  \includegraphics[width=1.1cm,angle=270,clip]{DATA/Ri1/plotyzu.png}
%  (e)
%  \includegraphics[width=1.1cm,angle=270,clip]{DATA/Ri1/plotyzv.png}
%  (f)
%  \includegraphics[width=1.1cm,angle=270,clip]{DATA/Ri1/plotyzT.png}
%  \\ \vskip1.em
%  (g)
%  \includegraphics[width=1.1cm,angle=270,clip]{DATA/Ri1_Ra7/plotyzu.png}
%  (h)
%  \includegraphics[width=1.1cm,angle=270,clip]{DATA/Ri1_Ra7/plotyzv.png}
%  (i)
%  \includegraphics[width=1.1cm,angle=270,clip]{DATA/Ri1_Ra7/plotyzT.png}
%  \\ \vskip1.em
%  \caption{Simulations at $\Ri_b=1$: instantaneous visualizations of $u'$ (left column), $v'$ (middle column), $\theta'$ (right column) in cross-stream plane for flow case RUN\_Ra9\_Re4.5 (first row), RUN\_Ra8\_Re4 (second row), RUN\_Ra7\_Re3.5 (third row). 24 contour levels are shown for each variable in the range $\pm 3$ standard deviations from the mean value (negative values in blue and positive in red).}
%  \label{fig:Ri_yz}
%\end{figure}

It is important to note that, to a first approximation, the type of flow pattern is controlled 
by the bulk Richardson number. As an illustrative example, in figure~\ref{fig:Ri_xz_CL}
we show flow visualizations for three flow cases with $\Ri_b=1$, in decreasing order of $\Rey_b$ (and $\Ra$). 
A similar type of large--scale organization is recovered 
in all three cases, perhaps with larger meandering of the rollers at the higher $\Rey_b$,
at which finer scale organization of turbulence is obviously also observed.

\begin{figure}
 \centering
  \includegraphics[width=12.cm,clip]{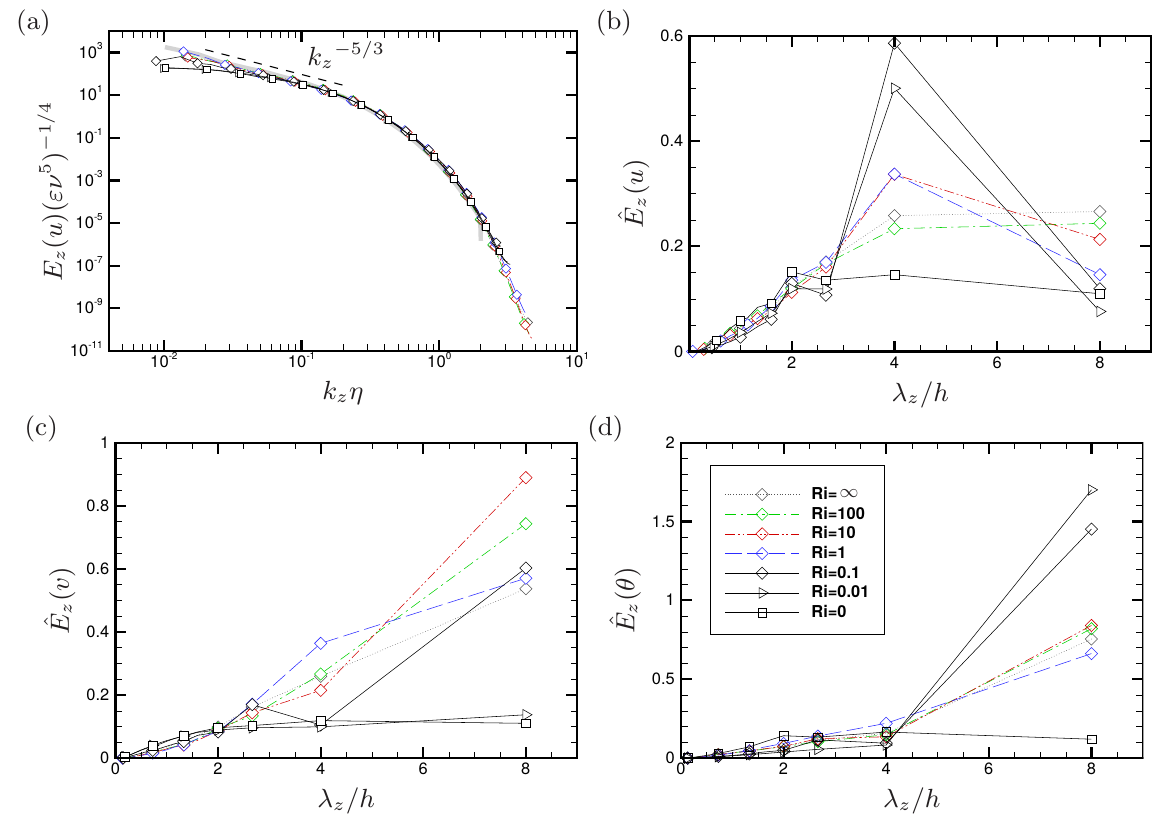}
  \caption{Spanwise spectral densities at channel centerline. In panel (a) the spectra of $v'$ are shown in Kolmogorov 
  representation (the thick grey line denotes the transverse velocity spectrum in isotropic turbulence at $\Rey_{\lambda}=142$~\citep{jimenez_93}).
  In panels (b), (c), (d) we show the normalized spectra of $u'$, $v'$, $\theta'$ as 
  a function of the wavelength. Refer to table~\ref{tab:nomen} for nomenclature of lines and symbols.}
 \label{fig:spectra_CL}
\end{figure}

More quantitative information regarding the flow structure can be gained by inspecting the spectral densities 
of the flow variables, providing information on the repartition of energy across the flow scales of turbulence.
The spanwise spectra are here considered as they are not affected by bulk
flow convection in the presence of shear~\citep{bernardini_14}.
In figure~\ref{fig:spectra_CL}(a) we show the spectra of wall--normal velocity fluctuations 
at the channel centerline in the classical
Kolmogorov representation for all the flow cases shown in the previous flow visualizations, 
hence spanning the entire range of Richardson numbers.
The figure shows near perfect universality of the distributions, and confirms adequate 
resolution of the small flow scales for all flow cases here reported. Excellent comparison is also obtained
with transverse velocity spectra obtained from DNS data for isotropic turbulence~\citep{jimenez_93},
hence supporting universality of the small scales far from walls.
To better highlight the different spatial organization at the large scales of motion, 
spectral densities of velocity components and temperature are shown in linear scale
and as a function of the spanwise wavelength in panels (b)--(d).
To rule out effects of large turbulence intensity variation with $\Rey_b$ and $\Ra$, 
the spectral densities are reported in normalized form (denoted with the hat symbol), 
in such a way that they all integrate to unity.
The figure confirms the dominance of energetic motions spanning the full channel width. 
In fact, the most energetic Fourier mode for $v$ and $\theta$ is found for all cases 
(with the exception of pure Poiseuille flow)
at $\lambda_z = L_z$, which is consistent with the previously noticed occurrence of two rollers 
in all flow visualizations.
It should be noted that the apparently non-monotonic Richardson number trend of the strength of the $L_z$ mode is 
due to the chosen normalization of the spectra. In fact, the unscaled spectra (not shown) have a monotonic
increasing trend as $\Ri_b$ decreases.
The behavior of the streamwise velocity spectra is quite different, and in that case the second Fourier 
mode ($\lambda_z = L_z/2$) is dominant, attaining a peak at $\Ri_b=0.1$. 
It is noteworthy that the doubled typical wavelength of $u$ with respect to $v$ is also different than observed in 
similar flows as plane Couette flow~\citep{pirozzoli_14}. The difference is due to the even symmetry properties of the mean velocity field 
with respect to the centerline, which implies that both upward and downward 
vertical motions locally convey positive streamwise velocity fluctuations. Hence, to each peak of $v'$ at the
channel centerline, two peaks of $u'$ must be present. 

\begin{figure}
 \centering
  \includegraphics[width=12.cm,clip]{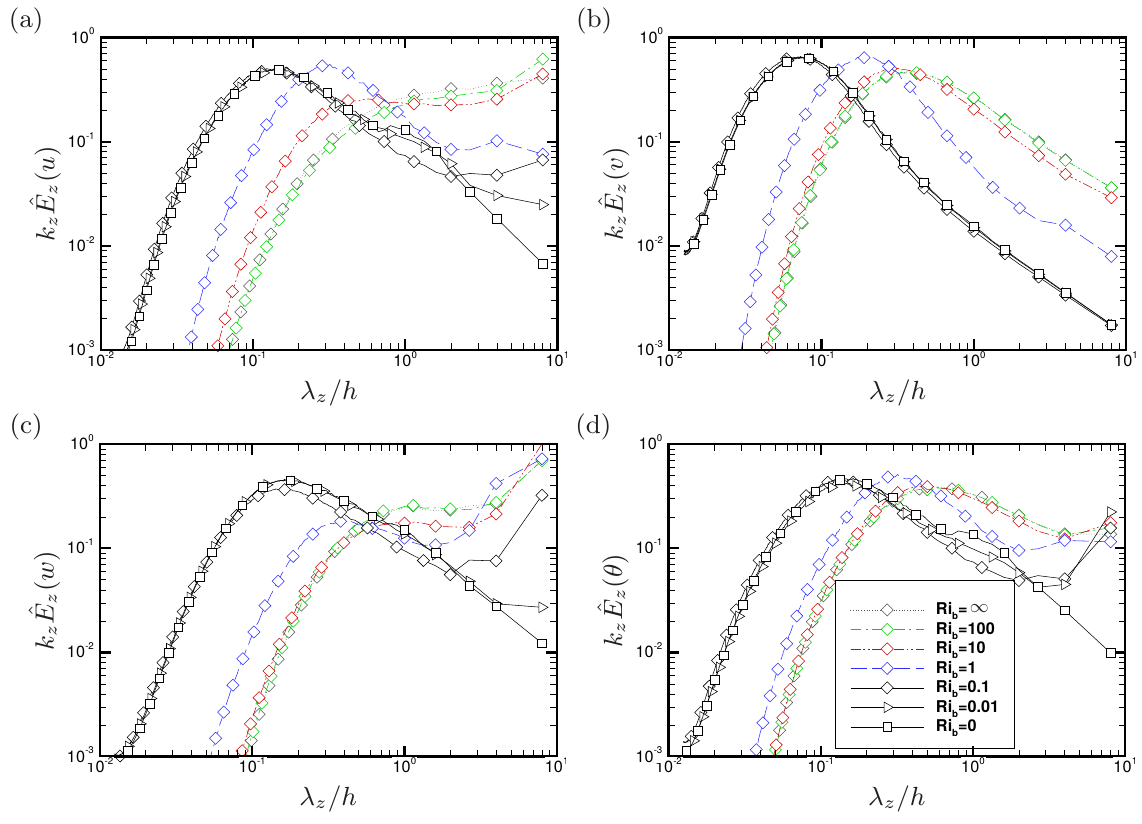}
  \caption{Pre-multiplied, normalized spanwise spectral densities of longitudinal velocity (a), vertical velocity (b), spanwise velocity (c) and temperature (c) at $y=y_P$.}
 \label{fig:spectra_NW}
\end{figure}

The spectral densities at the near--wall station ($y=y_P$) are shown in figure~\ref{fig:spectra_NW}.
In this case, a semi--logarithmic representation is used for the pre--multiplied spectra, in such a way
that equal areas correspond to equal energies.
In flow cases near the free--convection limit the spectra of $v$ and $\theta$ are bump--shaped, with a maximum
at $\lambda_x \approx 0.3 h$, which is connected with the typical spacing between adjacent
near--wall plumes. %, whose size is proportional to the thermal boundary layer thickness.
On the other hand, the $u$ and $w$ spectra feature energy concentration at the largest scales,
which is the footprint of the rollers sweeping the walls by horizontal `winds'
because of the impermeability condition.
A change of behavior is noticed around $\Ri_b=1$, which marks a substantial reduction 
of the typical length scale of the $v$-bearing eddies towards $\lambda_z \approx 0.07 h$, which
corresponds to about 50 wall units (see table~\ref{tab:params}). 
This is the typical scale found in the near-wall streaks in Poiseuille flow~\citep{kim_87},
hence this change of flow scales is the symptom of passage from the regime of free to forced convection.
A bump also forms in the spectra of $u$, $w$ and $\theta$ at $\Ri_b \lesssim 1$, 
corresponding to about 100 wall units, again consistent with the behavior in Poiseuille flow.
Notably, at intermediate Richardson numbers, the spectra of $w'$ seem to contain more energy at the largest resolved 
modes than the spectra of $u'$, which can be explained recalling the dominant streamwise alignment of the rollers 
in the intermediate $\Ri_b$ regime.

%\begin{figure}
% \centering
%  \psfrag{x}[t][][1.0]{$\lambda_z/h$}
%  (a)
%  \psfrag{y}[b][][1.0]{$\hat{E}_z(u)$}
%  \includegraphics[width=4.0cm,clip]{DATA/spectra_u_CL_log_Ri1.eps}
%  (b)
%  \psfrag{y}[b][][1.0]{$\hat{E}_z(v)$}
%  \includegraphics[width=4.0cm,clip]{DATA/spectra_v_CL_log_Ri1.eps} \\
%  (c)
%  \psfrag{y}[b][][1.0]{$\hat{E}_z(\theta)$}
%  \includegraphics[width=4.0cm,clip]{DATA/spectra_T_CL_log_Ri1.eps}
%  \caption{Simulations at constant $\Ri_b$: normalized spanwise spectral densities of longitudinal velocity (a), vertical velocity (b) and temperature (c) at channel centerline.}
% \label{fig:spectra_CL_Ri1}
%\end{figure}

%\begin{figure}
% \centering
%  \psfrag{x}[t][][1.0]{$\lambda_z$}
%  (a)
%  \psfrag{y}[b][][1.0]{$k_z \hat{E}_z(u)$}
%  \includegraphics[width=4.0cm,clip]{DATA/spectra_u_NW_Ri1.eps}
%  (b)
%  \psfrag{y}[b][][1.0]{$k_z \hat{E}_z(v)$}
%  \includegraphics[width=4.0cm,clip]{DATA/spectra_v_NW_Ri1.eps} \\
%  (c)
%  \psfrag{y}[b][][1.0]{$k_z \hat{E}_z(\theta)$}
%  \includegraphics[width=4.0cm,clip]{DATA/spectra_T_NW_Ri1.eps}
%  \caption{Simulations at constant $\Ri_b$: pre-multiplied normalized spanwise spectral densities of longitudinal velocity (a), vertical velocity (b) and temperature (c) at $y=y_P$.}
% \label{fig:spectra_NW_Ri1}
%\end{figure}

\section{Flow statistics} \label{sec:stats}

\begin{figure}
 \centering
  \includegraphics[width=12.cm,clip]{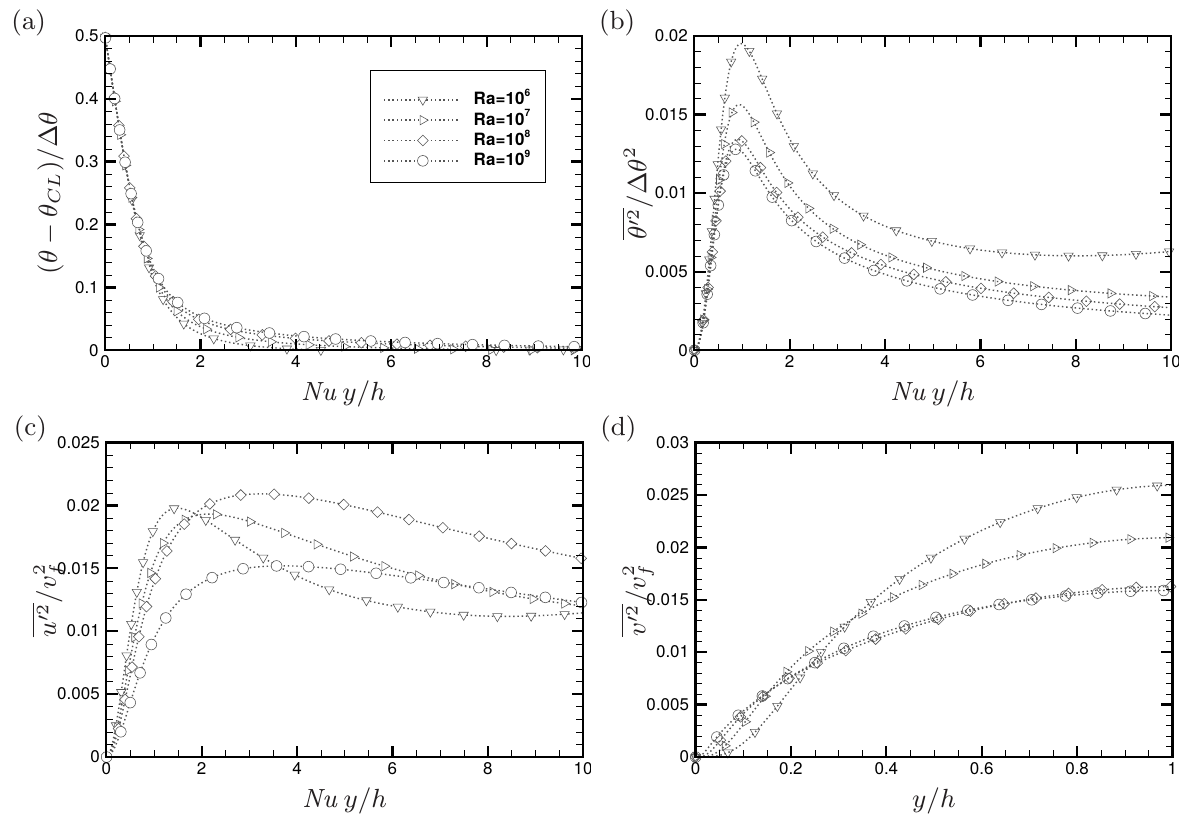}
  \caption{Profiles of mean temperature (a), temperature variance (b), 
           horizontal velocity variance (c), and vertical velocity variance (d) 
           in pure Rayleigh-B\'enard flow ($\Ri_b = \infty$). $v_f = (2 \beta g h \Delta \theta)^{1/2}$ is the
           reference free-fall velocity, and the subscript CL is used to indicate the channel centerline state.}
 \label{fig:tmean_RB}
\end{figure}

The main flow statistics are presented in this section, starting from the limiting cases of pure 
free and forced convection. The results obtained for Rayleigh--B\'enard convection are
shown at several Rayleigh numbers in figure~\ref{fig:tmean_RB}, with temperatures scaled by the 
total difference $\Delta \theta$, and velocities scaled with the reference free-fall velocity 
$v_f = (2 \beta g h \Delta \theta)^{1/2}$. For the sake of comparison of statistics at different Rayleigh numbers,
wall distances (with the exception of panel (d)) are multiplied by the respective Nusselt number, as $h/\Nu$ is proportional
to the thermal boundary layer thickness~\citep{ahlers_09}. In fact, the mean temperature profiles
in panel (a) show near collapse in this representation, with an extended nearly linear profile, 
consistent with the established motion that at the (relatively low) Rayleigh numbers under scrutiny 
the boundary layer is in a (quasi--)laminar state~\citep{ahlers_09}. The location $y=h/\Nu$ also very well matches
the location where temperature fluctuations have a maximum (panel (b)), and the peak location of 
$\theta'$ production (namely $y_P$). It is noteworthy that
the amplitude of this maximum depends on the Rayleigh number to some extent, with possible saturation at sufficiently high $\Ra$.
This effect is likely caused by the decreased amplitude 
of vertical motions when measured in $v_f$ units (panel (d)), which also show evidence for saturation at high $\Ra$. Apparently, saturation of $\theta'$ in confined geometries is not observed at Reyleigh numbers as high as $10^{11}$~\citep{stevens_11}.
Similar observations were made by \citet{orlandi_15}, who noticed that in pure shear flow turbulent fluctuations (as compared to the bulk channel velocity)
are in fact higher at lower Reynolds number. No systematic trend with $\Ra$ is observed for the horizontal velocity 
fluctuations (panel (c)), which are likely dominated by large--scale sweeping motions.

The structure of the velocity and temperature fields in natural convection has been the subject of 
extensive speculations within the meteorological community~\citep{wyngaard_92}.
It is a common notion that a free--convection layer exists in the atmosphere in which
wind and temperature exhibit inverse power--law scalings with the wall distance
\begin{equation}
\begin{array}{ccc}
\displaystyle \frac {\overline{\theta}(y) - \theta_0}{\theta_{\tau}} &=& 3 B_{\theta} \left( (y/L)^{-1/3} - (y_0/L)^{-1/3} \right), \\
\displaystyle \frac {\overline{u}(y) - u_0}{u_{\tau}} &=& - 3 B_{u} \left( (y/L)^{-1/3} - (y_0/L)^{-1/3} \right), 
\label{eq:Prandtl}
\end{array}
\end{equation}
where $\theta_{\tau}=Q/u_{\tau}$ is the friction temperature, $L$ is the Monin-Obukhov length scale defined in equation~\eqref{eq:MO},
the subscript $0$ denotes a suitable off--wall reference location, and $B_{\theta}$ and $B_u$ 
are two supposedly universal constants (\citet{kader_90} report $B_{\theta} \approx 1.3$, $B_u \approx 1.9$).
Equation~\eqref{eq:Prandtl} was first derived by \citet{Prandtl_32} from mixing length arguments,
based on the assumption that the typical vertical velocity scale of buoyant plumes is $v_P = (\beta g Q y)^{1/3}$,
the associated temperature scale is $\theta_P = Q^{2/3} (\beta g y)^{-1/3}$, and the turbulent Prandtl number is constant.
Hence, the accompanying expectation for velocity fluctuations is that they grow as $y^{1/3}$, whereas 
temperature fluctuations should decay as $y^{-1/3}$.
\begin{figure}
 \centering
  \includegraphics[width=12.cm,clip]{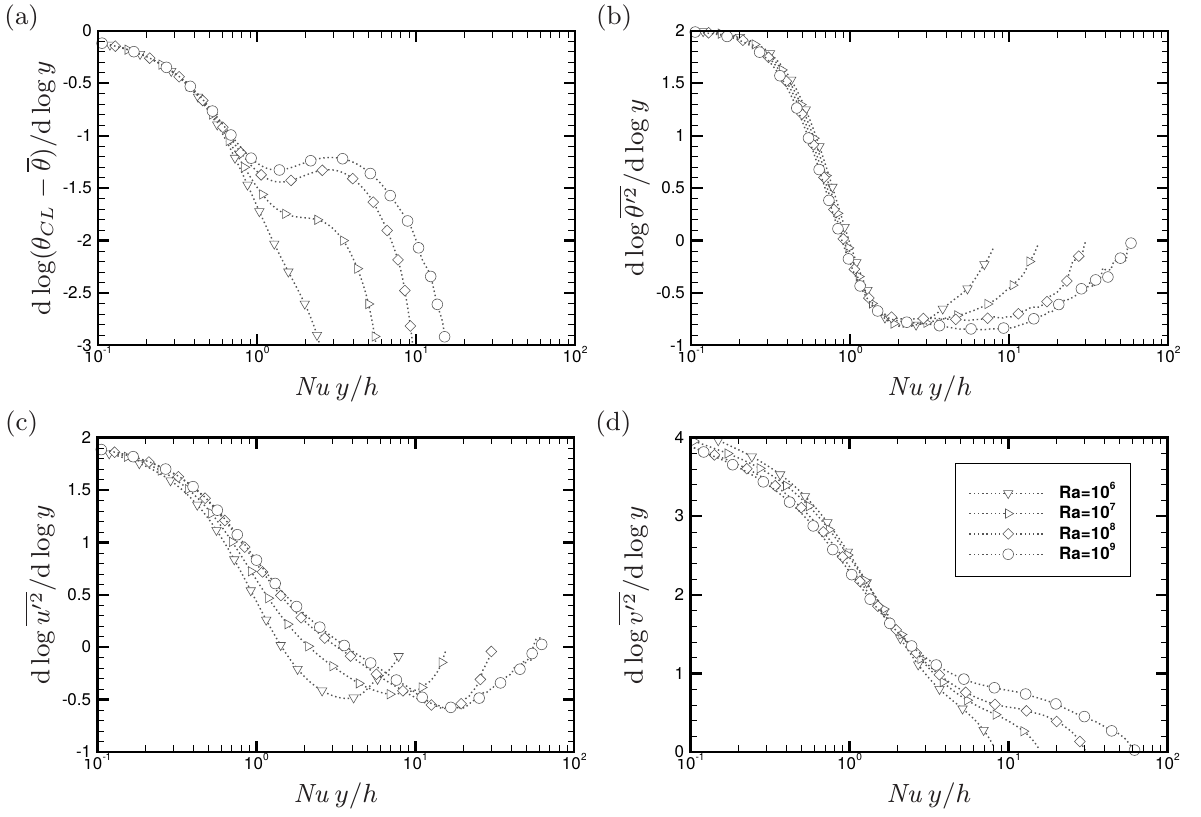}
  \caption{Power-law diagnostic functions for mean temperature (a) temperature variance (b),
           horizontal velocity variance (c), and vertical velocity variance (d), in Rayleigh-B\'enard flow ($\Ri_b = \infty$).}
 \label{fig:diag_RB}
\end{figure}
Atmospheric measurements~\citep{kader_90} suggest that the scaling laws~\eqref{eq:Prandtl}
are grossly satisfied in the free--convection regime. However, field experiments in that regime 
are inevitably affected by the presence of (albeit small) mean winds, which implies
large uncertainties and lack of reproducibility. On the other hand, laboratory experiments  
of pure convection have mostly focused on strongly confined conditions~\citep{ahlers_09},
hence do not convey much useful information for the purpose.
To directly verify the occurrence of the inverse power--law behaviour given by~\eqref{eq:Prandtl} 
in our DNS data we consider the distributions of the power--law indicator functions, defined as 
$\Xi_{\varphi} = (y/\varphi) \, \diff \varphi/ \diff y$ for several flow variables.
The presence of a plateau in $\Xi_{\varphi}$ would obviously indicate a $\varphi \sim y^{\Xi_{\varphi}}$
power--law range.
The power--law indicators for mean temperature and for velocity and temperature variances 
are shown in figure~\ref{fig:diag_RB} for free convection flow cases.
The mean temperature (panel (a)) does not show any evidence of a sensible layer with power--law behaviour.
It may be argued that a dip is forming in a narrow range of wall distances at the highest Rayleigh number here considered ($\Ra=10^9$),
although the resulting power--law exponent is far from the alleged $-1/3$. This large
discrepancy with respect to the Prandtl's theory seriously puts into question the validity of the predicted scaling,
or perhaps suggests that extreme Rayleigh numbers are required to observe the $y^{-1/3}$ law.
On the other hand, it is found that an extended power--law region forms for the temperature fluctuations
with exponent not far from the expected $-2/3$ value, and which widens with $\Ra$. The situation is even less clear for the velocity fluctuations.
Specifically, while the variance of vertical fluctuations may seem to form a small plateau with positive power--law exponent
(optimistically, not far from the expected $2/3$), streamwise fluctuations are consistently decreasing
towards the channel centerline, thus clearly contradicting the expected increasing trend.
This odd behavior of velocity fluctuations in free convection has been long recognized, and deviations 
from Prandtl's scaling have been frequently attributed to the important effect of $h$--scaled eddies
on the horizontal velocity components~\citep{panofsky_77}.
\citet{kader_90} also predicted a $y^{-2/3}$ scaling for the horizontal 
velocity variances, based on the assumption that the relevant velocity scale for 
horizontal velocity fluctuations is $u_{\tau}^2/v_P$. 
Our impression is that figure~\ref{fig:diag_RB}(c) at least partially confirms their prediction,
although no extended power--law range is observed here.

\begin{figure}
 \centering
  \includegraphics[width=12.cm,clip]{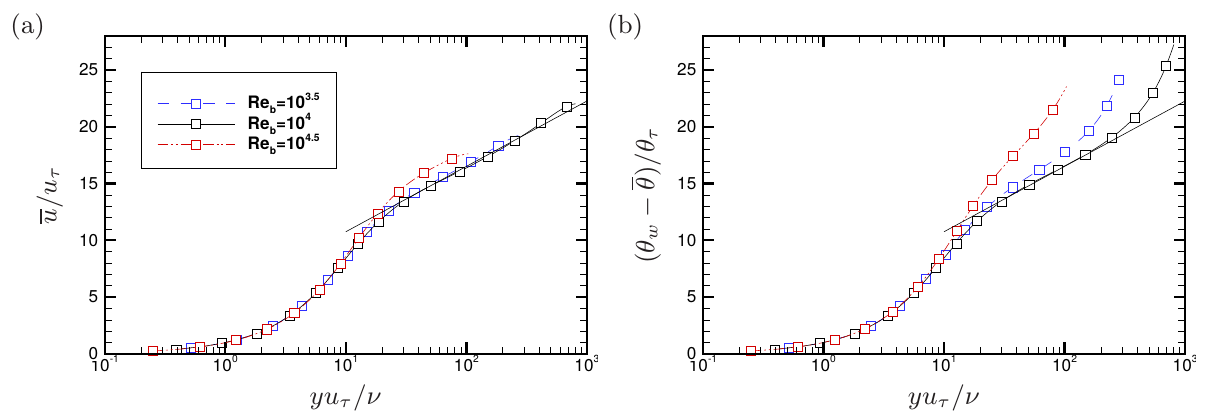}
  \caption{Profiles of mean streamwise velocity (a) and mean temperature (b) for pure Poiseuille flow ($\Ri = 0$).
           The solid lines correspond to logarithmic laws for velocity and temperature.}
 \label{fig:umean_PO}
\end{figure}

In the sake of completeness the mean velocity and temperature profiles for the case
of pure Poiseuille flow are also shown in figure~\ref{fig:umean_PO}. 
It is found that the mean velocity profiles (panel (a)) are accurately fitted using the 
conventional logarithmic approximation, namely $u/u_{\tau} = C + \log (y u_{\tau} / \nu) / k$,
with the traditional choice $C=5$, $k=0.4$. The same approximation works well also
for the temperature field (panel (b)), hence at unit Prandtl number and in the absence of coupling through buoyancy, 
very close similarity if found between the velocity and the temperature fields, except in the channel core, where temperature 
must have non--zero derivative.
%A detailed study of the behavior of passive scalars in Poiseuille flow up to 
%$\Rey_{\tau} \approx 4000$ has been recently carried out by \citet{pirozzoli_16}.

\begin{figure}
 \centering
  \includegraphics[width=12.cm,clip]{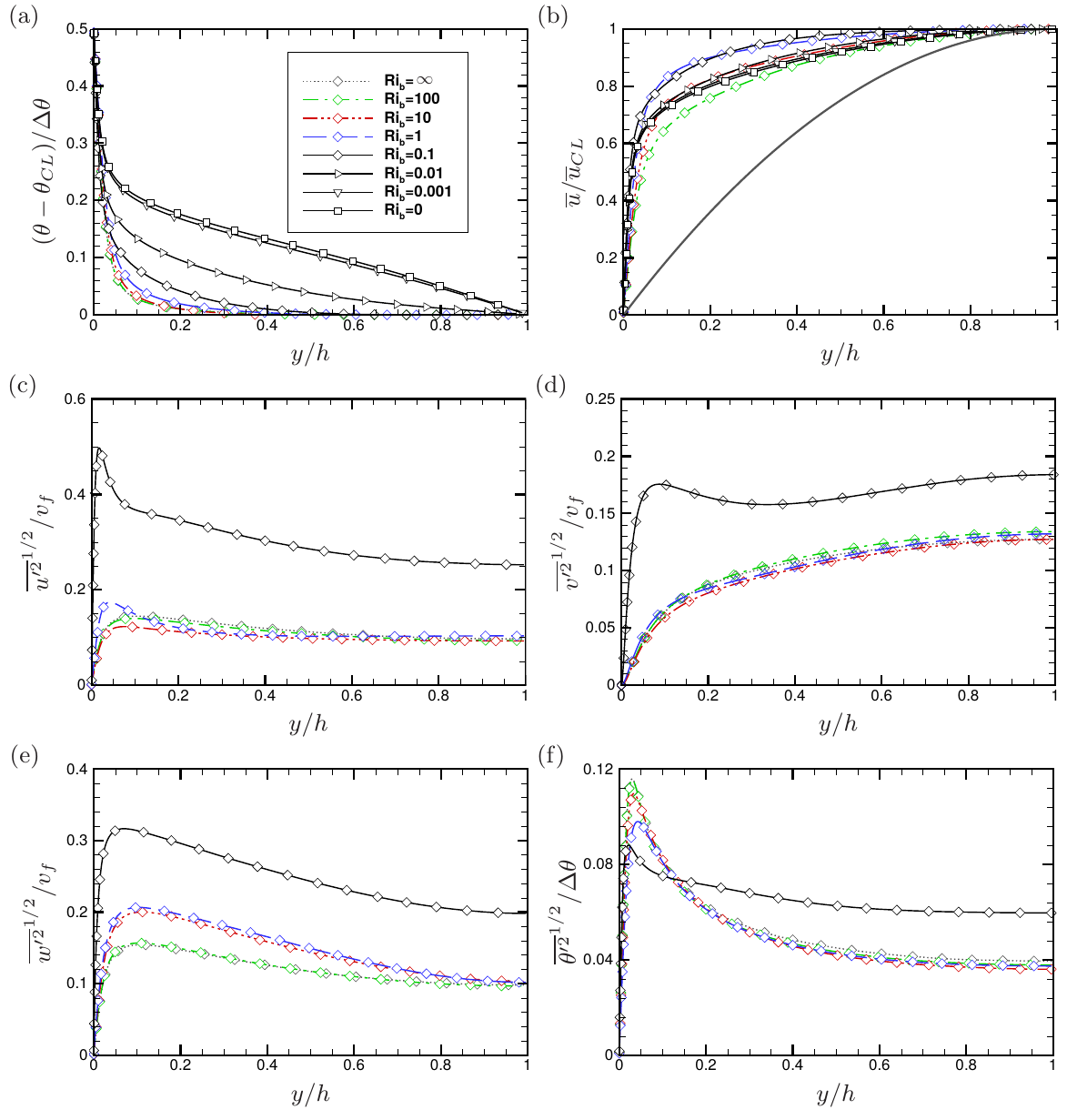}
  \caption{Profiles of mean temperature (a), mean streamwise velocity (b), 
           streamwise velocity variance (c), vertical velocity variance (d), 
           spanwise velocity variance (e) and temperature variance (f).
%          in near R-B regime ($\Ri \ge 0.1$). 
           The thick grey lines in panel (b) corresponds to the laminar Poiseuille parabolic profile.}
 \label{fig:tmean_vs_Ri}
\end{figure}

The effect of Richardson number variation starting from the free--convection regime is illustrated
in figure~\ref{fig:tmean_vs_Ri}. As $\Ri_b$ decreases (i.e. the mass flow rate increases at given $\Ra$) 
the mean temperature profile (a) tends to become less flat, departing from the free--convection distribution,
and eventually attaining a near logarithmic distribution.
The mean velocity profile (b) has a more complex behavior, initially becoming more blunted down to $\Ri_b = 0.1$, 
and then less flat while approaching the Poiseuille limit.
Although the bulk Reynolds number is very low in the light--wind (high-$\Ra$) simulations, the velocity profile seems to be
much different than the laminar Poiseuille profile, which is also shown in panel (b) for reference. 
For instance, the flow case RUN\_Ra8\_Re3 (corresponding to $\Ri_b=100$) shows a blunted profile despite having a
bulk Reynolds number well below the expected threshold for forced turbulence to sustain itself in pure shear flow.
This intermediate state is not found in either of the limit cases of 
Rayleigh-B{\'e}nard and Poiseuille flow, while having some features of both.
%SERGIO questa frase (quella sopra) mi sembra piuttosto vaga 
Regarding the statistics of turbulent fluctuations (only shown here for $\Ri_b \ge 0.1$ for clarity), 
a very similar pattern as in free convection is observed down to $\Ri_b = 1$.
A different regime starts at $\Ri_b = 0.1$, which is associated with increased importance of shear. Hence, near--wall peaks 
of $u'$ and $v'$ form, and the profile of $\theta'$ tends to flatten in the channel core.

\begin{figure}
 \centering
  \includegraphics[width=12.cm,clip]{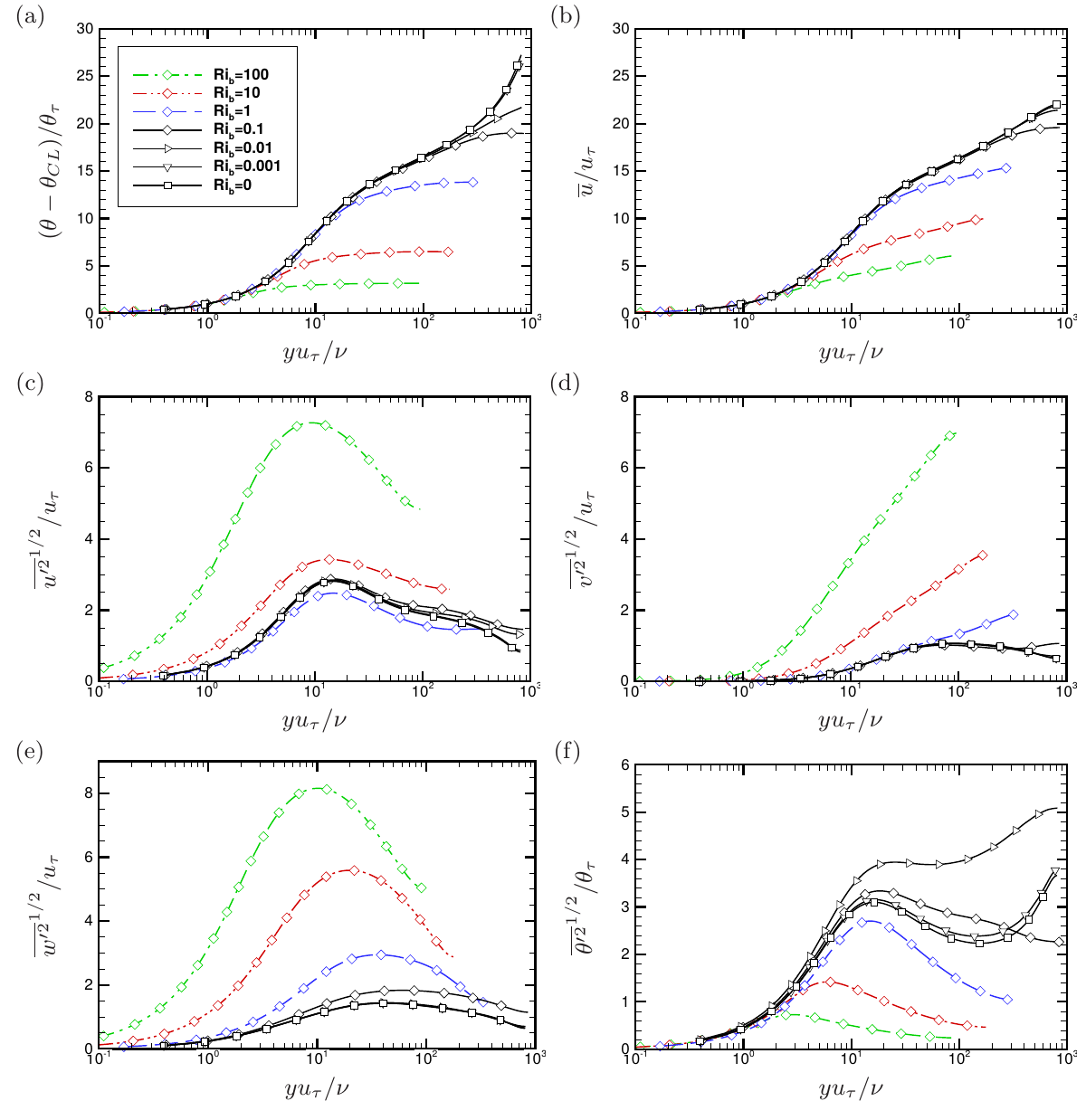}
  \caption{Shear-scaled profiles of mean temperature (a), mean streamwise velocity (b),
           streamwise velocity variance (c), vertical velocity variance (d), 
           spanwise velocity variance (e) and temperature variance (f),
           in mixed convection regime.}
 \label{fig:tmean_vs_Ri_inner}
\end{figure}

\begin{figure}
 \centering
  \includegraphics[width=12.cm,clip]{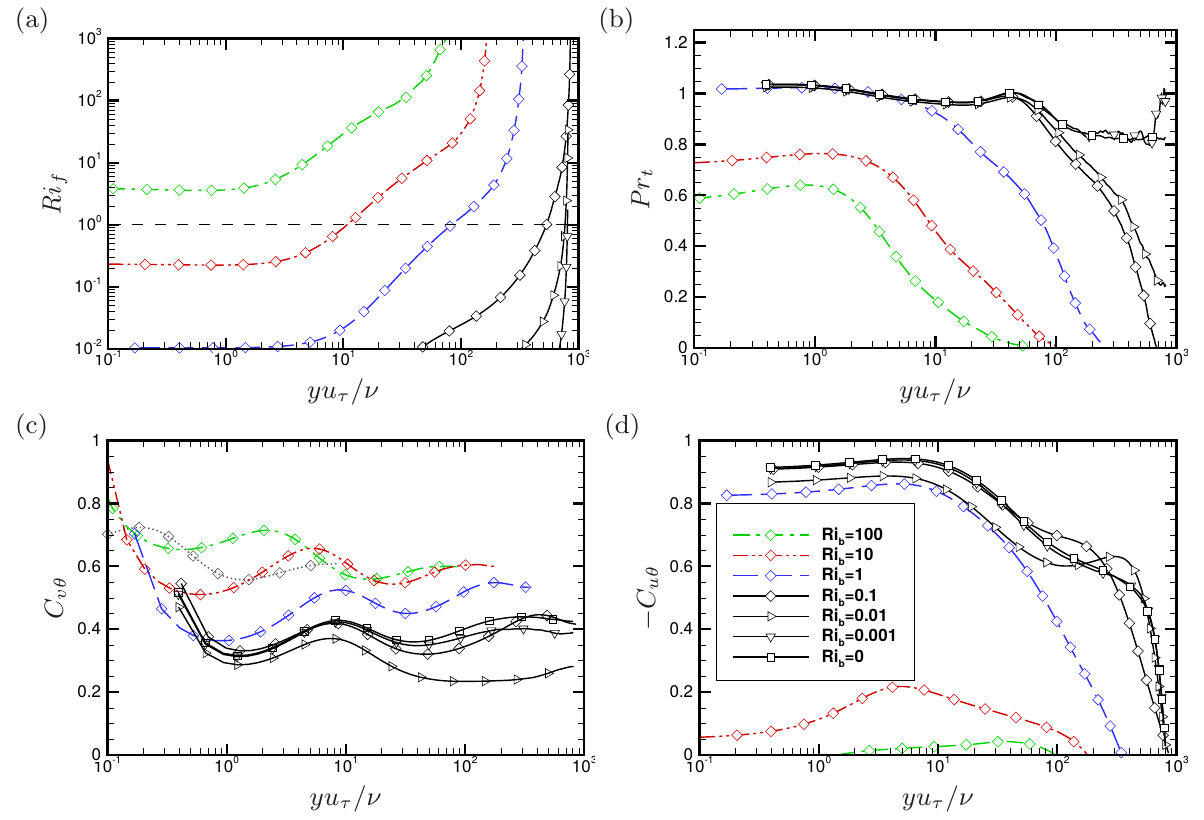}
  \caption{Profiles of flux Richardson number (a), turbulent Prandtl number (b),
           $v-\theta$ correlation coefficient (c) and $u-\theta$ correlation coefficient in the mixed convection regime.
           The dashed horizontal line in panel (a) marks the unit value of $\Ri_f$.}
 \label{fig:Rif_vs_Ri_inner}
\end{figure}

The same properties are reported upon wall scaling in figure~\ref{fig:tmean_vs_Ri_inner},
to highlight deviations from the forced convection limit at increasing $\Ri_b$.
In wall units, both the temperature and the velocity profiles approach the zero--$\Ri_b$ log--law limit
from below, and logarithmic layers for both variables are observed starting at $\Ri_b = 0.1$.
A universal wall scaling for the velocity and temperature fluctuations is also 
established at $\Ri_b = 0.1$, with near--wall peaks of $u'$ and $\theta'$ at $y^+ \approx 15$, and 
peaks of $v'$ and $w'$ further away. At higher $\Ri_b$ increasing values are found for
the wall---scaled velocity fluctuations, and lower values are found for the temperature fluctuations. 
This is the result of a change from wall scaling to free---fall scaling, since
here $\Rey_b$ is decreasing at constant $\Ra$, and as a result $v_f / u_{\tau} = \sqrt{\Ra/\Pran}/(2 \Rey_{\tau}$ 
is increasing, and $\theta_{\tau}/\Delta \theta$ is decreasing at increasing $\Ri_b$.
Furthermore, the peaks of $u'$ and $w'$ tend to come close to each other because of 
flow isotropization in wall---parallel planes.
%DAQUI

Further insights into changes occurring between $\Ri_b=0.1$ and $1$ can be
gained from figure~\ref{fig:Rif_vs_Ri_inner}. In panel (a) we report the distributions 
of the flux Richardson number
\begin{equation}
\Ri_f = \frac{-\beta g \overline{v' \theta'}}{\overline{u'v'} \diff \overline{u} / \diff y}, \label{eq:Rif}
\end{equation}
representing the ratio of the production of vertical velocity variance to the production of 
horizontal velocity variance. Hence, $\Ri_f$ is expected to be a local indicator of the relative
dynamical importance of buoyancy as compared to shear.
Except for the limiting case of high $\Ri_b$, the near--wall region is always dominated 
by shear, hence it is referred to as dynamic or convective sublayer~\citep{kader_90} in the meteorological community.
Further up, the flow in the so--called free--convection layer is dominated by buoyancy.
A major change occurs between $\Ri_b=1$, where the dynamic sublayer occupies about $10\%$ of the wall layer,
and $\Ri_b=0.1$, where this fraction exceeds $50\%$.
A quantity of great relevance in turbulence models for scalar transport is the
turbulent Prandtl number, defined as the ratio of the turbulent momentum and temperature diffusivities, namely
\begin{equation}
\Pran_t = \frac{\nu_t}{\alpha_t} = \frac {\overline{u' v'}}{\overline{v' \theta'}} \frac {\diff \overline{\theta} / \diff y}{\diff \overline{u} / \diff y}, \label{eq:prt}
\end{equation}
whose distribution is shown in figure~\ref{fig:Rif_vs_Ri_inner}(b). 
Consistently with numerical and experimental data~\citep{cebeci_84,kader_81,pirozzoli_16}, 
in the forced convection regime $\Pran_t$ is close to unity in the near--wall region, and to $0.85$ 
in the bulk flow up to $y/h \approx 0.5$.
Notably, the Prandtl number starts dropping from the outer layer at $\Ri_b = 0.01$, 
and its value is well below unity throughout the channel at high $\Ri_b$.
This behaviour is in clear contradiction with the assumtion of constant $\Pran_t$ advocated in Prandtl's free--fall theory,
and it clearly indicates that buoyancy is much more effective in redistributing temperature than momentum.
This behaviour is confirmed by the $v-\theta$ and $u-\theta$ correlation
coefficients, shown in panels (c) and (d), respectively.
As found in canonical Poseuille flow~\citep{pirozzoli_16}, $C_{v\theta}$ stays close
to $0.4$ throughout the wall layer at low $\Ri_b$, and it increases with $\Ri_b$ reaching a value 
of about $0.6$ in free convection, reflecting increased effectiveness of wall--normal motions.
On the other hand, $C_{u\theta}$ (obviously negative) is found to be 
close to $0.9$ near the wall and to decrease with the wall distance in forced convection,
reflecting the similarity in the behavior of $u$ with that of a passive scalar~\citep{abe_09}.
The correlation drops sharply past $\Ri_b = 1$, reflecting the change in the organization of $u$
by buoyancy--induced vertical motions, presumably through pressure effects.

\section{Monin-Obukhov scaling} \label{sec:MO}

\begin{figure}
 \centering
  \includegraphics[width=12.cm,clip]{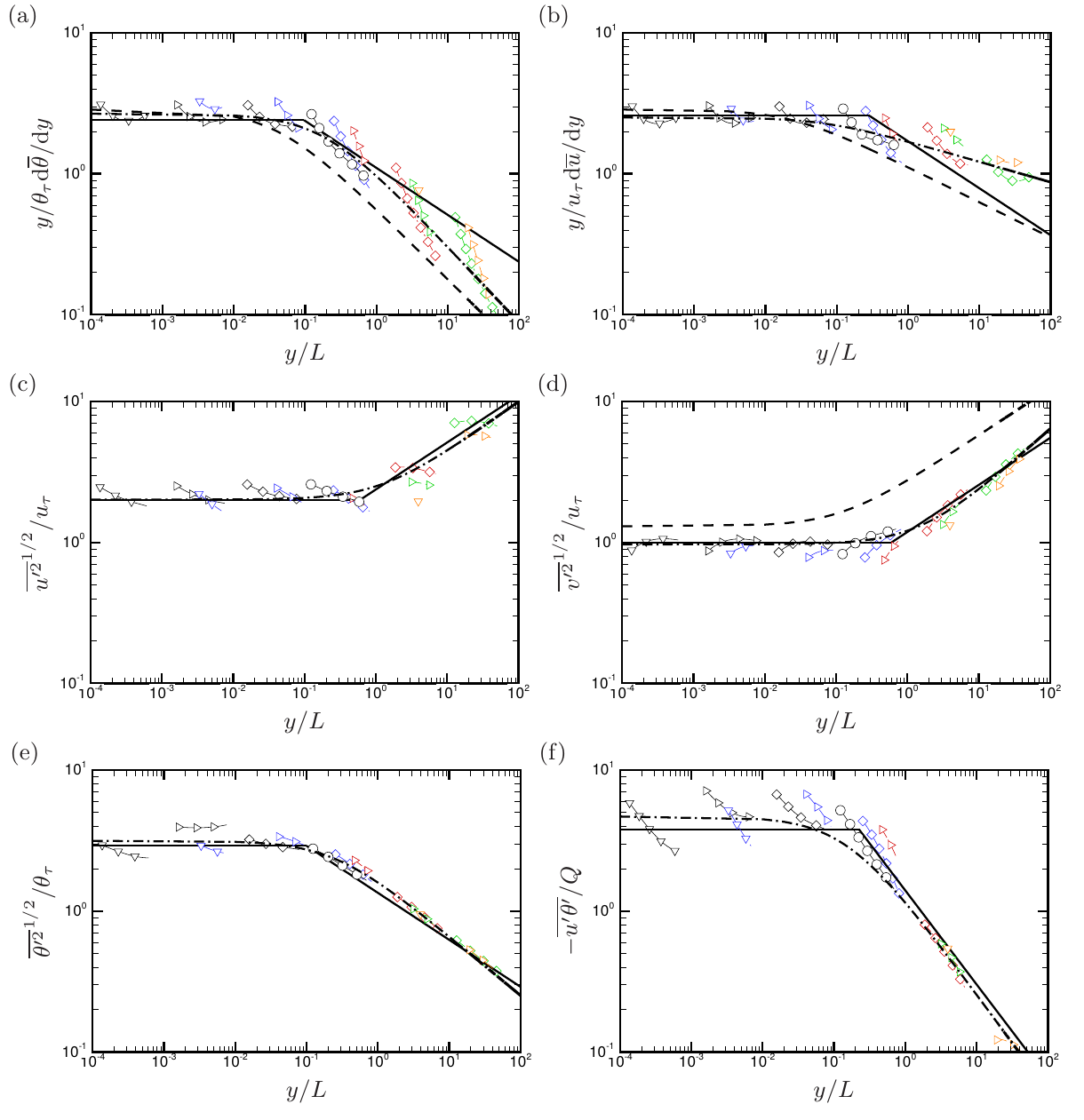}
  \caption{Verification of Monin-Obukhov similarity hypothesis for: (a) mean temperature gradient; (b) mean velocity gradient; (c) streamwise velocity variance; (d) vertical velocity variance; (e) temperature variance; (f) $u-\theta$ correlation. The solid lines indicate a compound of the scaling laws proposed by \citet{kader_90}, the dashed lines denote the Businger-Dyer relationships with classical values of the constants, whereas the dot-dashed lines indicate a fit of the DNS data (see table~\ref{tab:BD}). See table~\ref{tab:nomen} for nomenclature of lines and symbols.}
 \label{fig:MO}
\end{figure}

A useful parametrization of the wall region in the presence of mixed convection is provided
by the Monin-Obukhov theory~\citep{obukhov_46,monin_54}. Starting from the assumption that the correct velocity
scale in wall bounded flows is the friction velocity $u_{\tau}$, and the only dimensionally correct length
scale is $L$ as defined in equation~\eqref{eq:MO}, the following scalings result for the 
fully turbulent part of the wall layer
\begin{eqnarray}
\frac{y}{\theta_{\tau}} \frac{\diff \overline{\theta}}{\diff y} = \varphi_{h} \left( \frac yL \right), \quad
\frac{y}{u_{\tau}} \frac{\diff \overline{u}}{\diff y} = \varphi_{m} \left( \frac yL \right), \\
\frac{\overline{u_i'^2}^{1/2}}{u_{\tau}} = \varphi_i \left( \frac yL \right), 
\frac{\overline{\theta'^2}^{1/2}}{\theta_{\tau}} = \varphi_{\theta} \left( \frac yL \right), \quad
\frac{-\overline{u'\theta'}}{Q} = \varphi_{u\theta} \left( \frac yL \right), \quad
\end{eqnarray}
 with $-\overline{u'v'} \approx \tau_w$, $-\overline{v'\theta'} \approx Q$ and
the $\varphi$'s a suitable set of universal functions.
The Monin--Obukhov relations are widely used in the meteorological practice and 
as wall functions in numerical simulations of atmospheric circulation
as they allow to estimate momentum and temperature fluxes from mean flow gradients 
evaluated away from the wall~\citep{deardorff_70,stull_12}. Regarding the choice of the universal functions,
the typical approach~\citep{kader_90} consists of interpolating 
between the extreme conditions of forced and free convection.
For instance, regarding the scaling of temperature, it is expected that in forced convection
a logarithmic layer of $\overline{\theta}$ forms, hence $\varphi_{h} \approx k_{\theta}$,
where $k_{\theta}$ is the Karman constant for passive scalars~\citep{kader_81}.
On the other hand, granted the validity of Prandtl's theory of free convection (as from equation~\eqref{eq:Prandtl}),
the scaling $\varphi_{h} \sim (y/L)^{-1/3}$ would result.
A frequently used representation for the universal functions of Monin--Obukhov theory 
consists of the Businger--Dyer relationships~\citep{businger_71,dyer_74},
which assume 
\begin{equation}
\varphi_h = \frac 1{k_{\theta}} \left( 1 + \gamma_h y/L \right)^{\alpha_h}, \quad
\varphi_m = \frac 1{k} \left( 1 + \gamma_m y/L \right)^{\alpha_m}, \quad
\end{equation}
with the typical choice of constants~\citep{paulson_70} $k = k_{\theta} = 0.35$, $\gamma_m = \gamma_h = 16$, $\alpha_h=-1/2$, $\alpha_m=-1/4$.
Hence, these relationships account for strong deviations of the mean temperature and velocity fields from the 
alleged $(y/L)^{-1/3}$ behavior in the limit of light winds.
Similar empirical relations have been proposed for the vertical velocity variances by \citet{panofsky_77},
which are also included in the figure.
In order to check the validity of the Monin--Obukhov similarity predictions, the DNS data are reported
in scaled form in figure~\ref{fig:MO}. For that purpose, data have been collected in a limited 
part of the wall layer, identified under the somehow arbitrary conditions that: 
i) the turbulent heat flux is higher than $90\%$ of the 
total flux; ii) the turbulent momentum flux is higher than $90\%$ of its maximum value. These conditions 
basically identify the assumptions made by the Monin--Obukhov theory that the viscous fluxes are 
negligible, and the total stress is approximately constant.
In figure~\ref{fig:MO} we also show the asymptotic trends suggested by \citet{kader_90}, as well as
the results obtained by fitting the DNS data with Businger--Dyer--like distributions,
with coefficients given in table~\ref{tab:BD}.
Panel (a) of figure~\ref{fig:MO} shows flat behaviour of the scaled temperature gradient
up to $y/L \approx 0.1$, followed by a global roll--off with power--law exponent close to the $-1/2$ 
value given by the Businger--Dyer relationships, but sensibly steeper than the $-1/3$ value 
expected in free convection. It should also be noted that the individual profiles are far from following 
the predicted scalings, but rather tend to follow much sharper inverse power laws, 
as previously discussed in figure~\ref{fig:diag_RB}(a). 
\begin{table}
 \centering
 \begin{tabular*}{1.\textwidth}{@{\extracolsep{\fill}}cccrrc}
  \hline
  Quantity & $k$ & $\gamma$ & $\alpha$ & $\alpha_{KY}$ & $\alpha_{BDP}$ \\
  \hline
  $\varphi_{h}$       & $0.375$    & $5.67$   & $-0.538$ & -1/3  & -1/2 \\
  $\varphi_{m}$       & $0.399$    & $14.6$   & $-0.145$ & -1/3  & -1/4 \\
  $\varphi_{1}$       & $0.498$    & $0.830$  & $0.361$  &  1/3  &   /  \\
  $\varphi_{2}$       & $1.03$     & $0.638$  & $0.452$  &  1/3  &  1/3 \\
  $\varphi_{\theta}$  & $0.318$    & $4.08$   & $-0.420$ & -1/3  &   /  \\
  $\varphi_{u\theta}$ & $0.214$    & $6.70$   & $-0.690$ & -2/3  &   /  \\
  \hline
 \end{tabular*}
 \caption{Coefficients of DNS data fits reported in figure~\ref{fig:MO}, assuming 
          functional dependence of the type $\varphi = 1/k (1+\gamma y/L)^{\alpha}$.
          The coefficients $\alpha_{KY}$ denote the expected power-law scaling exponents 
          in the free-convection regime predicted by \citet{kader_90}, and the
          $\alpha_{BDP}$ those used in the classical Businger--Dyer and Panofsky relationships.}
 \label{tab:BD}
\end{table}
Similar observations have been occasionally made in the literature~\citep{khanna_97},
and attributed to the importance of $h$--scaled circulatory motions, which would imply
the inclusion of $h/L$ (commonly referred to as the stability parameter) as an additional parameter 
in the Monin--Obukhov functional relationships.
Incidentally we note that, based on the present DNS data, the stability parameter
is approximately a unique function of the bulk Richardson number, and we find $h/L \simeq 3.34 \Ri_b^{0.85}$.
The scaled mean velocity gradient (figure~\ref{fig:MO}(b)) also has a flat behavior 
up to $y/L \approx 0.1$, with very slow roll--off, probably slower that the $(y/L)^{-1/4}$ given by the
Businger--Dyer relationships, although the data scatter here is quite severe, and
again individual profiles follow steeper power laws.
As pointed out by \citet{rao_06}, the parametrization of the velocity field in the presence 
of light wind under unstable stratification is a weak point of weather forecast 
models, mainly because field experiments typically convey large scatter
for obvious difficulties in achieving stable flow conditions, 
For instance, based on field measurements, \citet{kader_90} even argue about possible 
inversion of the $\varphi_m$ curve at high $y/L$, which however finds no support in our data.
DNS is especially valuable as sustained flow conditions are achievable, although
higher values of the Reynolds number would be clearly desirable.
The velocity and temperature fluctuations (panels (c)--(e)) globally follow the Monin--Obukhov scalings 
quite well. Especially satisfactory is the behavior of the vertical velocity and temperature fluctuations
in the light--wind regime, as previously noticed regarding the free--convection convection (see figure~\ref{fig:diag_RB}b,d).
This finding probably points to the fact that vertical plumes are well parametrized by Monin-Obukhov theory,
whereas large--scale circulatory motions obey to a different scaling. In this respect, \citet{panofsky_77}
pointed out that the correct scale for horizontal velocity fluctuations
is probably the one defined by \citet{deardorff_70}, namely $v_D = (\beta g Q h)^{1/3}$,
which corresponds to Prandtl's free-fall scale based on the channel height.
Assuming $\overline{u'^2}^{1/2} \sim v_D$ implies $\overline{u'^2}^{1/2} / u_{\tau} \sim (h/L)^{1/3}$,
hence explaining why the global trend of the streamwise velocity fluctuations with $y/L$ is increasing,
whereas the individual profiles are decreasing with $y$ (also recalling the discussion made on figure~\ref{fig:diag_RB}c).
In this context it is quite surprising that the $u$--$\theta$ correlation (panel (e)) satisfies the expected
$-2/3$ scaling law in the free-convection regime~\citep{kader_90}, whereas the individual profiles have strong scatter under
conditions of near-neutral stratification. 

\section{Parametrization of heat transfer and skin friction} \label{sec:heat}

%\begin{figure}
% \centering
%  \psfrag{x}[t][][1.0]{$\Ra$}
%  (a)
%  \psfrag{y}[b][][1.0]{$\Nu$}
%  \includegraphics[width=4.5cm,clip]{DATA/Nu_vs_Ra_Re.eps}
%  (b)
%  \includegraphics[width=4.5cm,clip]{DATA/Nu_vs_Ra_Ri.eps} \\
%  \caption{Distributions of Nusselt number as a function of Rayleigh number, grouped into lines with same $\Rey_b$ (a) and with same $\Ri_b$ (b).
%           The solid and dashed lines in (b) denote a $\Ra^{0.3}$ and $\Ra^{1/2}$ trend, respectively.}
% \label{fig:Nu}
%\end{figure}

\begin{figure}
 \centering
  \includegraphics[width=12.cm,clip]{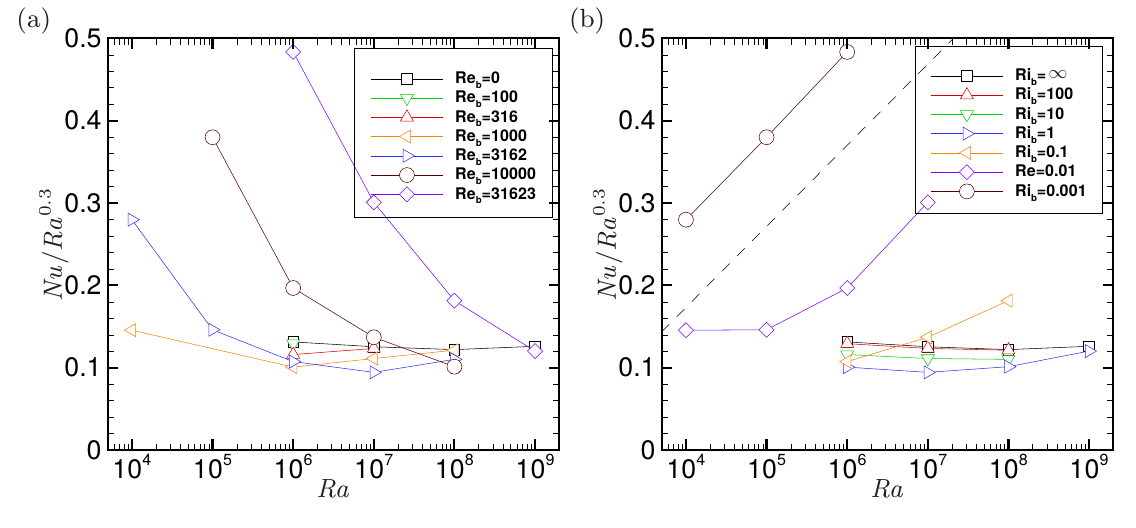}
  \caption{Compensated distributions of Nusselt number as a function of Rayleigh number, grouped into lines with same $\Rey_b$ (a) and with same $\Ri_b$ (b).
           The dashed line in panel (b) indicates a $\Ra^{0.45}$ power law.}
 \label{fig:Nu}
\end{figure}

\begin{figure}
 \centering
  \includegraphics[width=12.cm,clip]{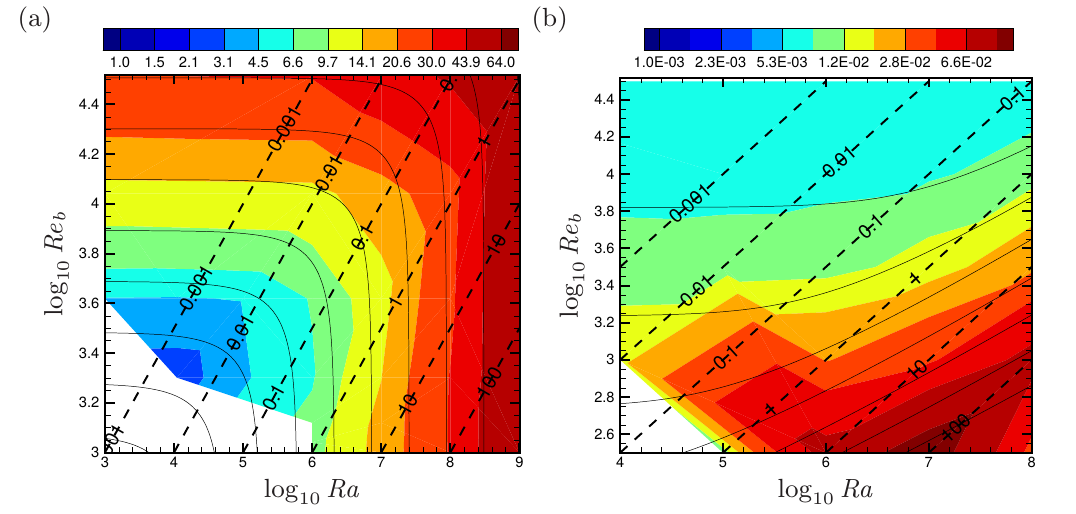}
  \caption{Maps of Nusselt number (a) and friction coefficient (b) as a function of Rayleigh and Reynolds number.
           The colored contours correspond to the DNS data, whereas the solid lines indicates fits
           obtained from equations~\eqref{eq:NuvsRaRe},\eqref{eq:Cffit}. The dashed diagonal lines
           have constant $\Ri_b$. Twelve logarithmically-spaced contour levels are shown for each variable,
           with $1 \le \Nu \le 64$, $0.001 \le C_f \le 0.1$.}
 \label{fig:NuCf2d}
\end{figure}

The prediction of heat transfer and aerodynamic drag under conditions of unstable stratification 
is a topic of obvious interest in engineering and meteorology. Given the absence of reliable theoretical 
mean temperature and velocity profiles for flow conditions far from neutral~\citep{scagliarini_15}, we attempt to derive
correlations based on the available DNS data. For that purpose, we preliminarily try to gain a perception for 
the behavior of heat transfer and friction coefficients as a function of the governing parameters.
In figure~\ref{fig:Nu} we show the computed Nusselt number, defined as
\begin{equation}
\Nu = \frac {2 Q h}{\alpha \Delta \theta}, \label{eq:Nu}
\end{equation}
as a function of the Rayleigh number, grouped into curves at constant Reynolds number (a) and constant Richardson number (b). Data are compensated by $\Ra^{0.3}$ to better highlight scatter among curves.
Panel (a) shows that data at sufficiently high $\Ra$ tend to cluster around
the free-convection ($\Ra=\infty$) distribution, with departures taking place at higher $\Ra$ as 
the Reynolds number increases, suggesting that the Richardson number may be an important
parameter to distinguish among different cases. The figure also shows 
that the Nusselt number is not a monotonic function of $\Rey_b$ for fixed $\Ra$, with the counterintuitive 
conclusion that some small amount of forced convection may yield reduction of the heat exchange 
with respect to the pure buoyant case, as also evident in panel (b).
This effect is the likely consequence of sweeping of convective plumes by large-scale motions,
with subsequent loss of efficiency in heat redistribution~\citep{scagliarini_14}.
Fitting the DNS data in the free--convection regime we obtain
\begin{equation}
\Nu(\Ra) \approx 0.1165 \, \Ra^{0.304}, \label{eq:NuvsRa}
\end{equation}
with a power--law exponent not too far from that typically reported
in this range of relatively low $\Ra$~\citep{ahlers_09,orlandi_15a}.
The effect of Richardson number reduction from pure buoyancy (see figure~\ref{fig:Nu}b) 
is initially a downward translation of
the $\Nu(\Ra)$ curve, with maximum reduction of up to $20 \%$ at $\Ri_b=1$.
Further reduction of $\Ri_b$ yields a marked increase in the slope of the $\Nu(\Ra)$ curve, 
which tends to attain a $\Ra^{0.45}$ slope in the forced convection limit.
%SERGIO: Roberto, questa frase inseriscila tu, please
%This is a likely consequence of the fact that the contribution of forced convection yields
%fully turbulent boundary layers, hence close to the alleged `ultimate state' of natural convection
%at which the theoretical trend is $\Ra^{0.5}$, except for logarithmic corrections~\citep{ahlers_09}.
%
%Se fossimo in regime di puro RB sarebbe .... In questo caso la stima però ,
%e questo è verosimilmente il motivo ...
%
%SERGIO, questo è un punto piuttosto spinoso perché la pendenza 1/2 è solo asintotica e a denominatore
%c'è una correzione [log(Ra)]^(3/2) che ai Ra delle nostre simulazioni porta ad un esponente effettivo
%di circa 0.38. Io direi che questa affermazione va molto "smoothata" per evitare discussioni infinite.
Fitting the DNS data herein reported in the $\Ri_b = 0$ limit
as well as data at higher $\Rey_b$~\citep{pirozzoli_16} yields
\begin{equation}
\Nu(\Rey_b) \approx 0.0073 \, \Rey_b^{0.802}. \label{eq:NuvsRe}
\end{equation}
%When comparing equation~\eqref{eq:NuvsRe} with classical formulas for forced convection care
%should be taken that the latter typically use the bulk temperature as reference.
A typical engineering approach~\citep{incropera_11} consists of using either equation~\eqref{eq:NuvsRa} 
if $\Ri_b \gtrsim 1$ or or equation~\eqref{eq:NuvsRe} if $\Ri_b \lesssim 1$.
A convex combination of the two formulas is also sometimes used
\begin{equation}
\Nu(\Ra,\Rey_b) = \left( \Nu(\Ra)^n + \Nu(\Rey_b)^n \right)^{1/n}, \label{eq:NuvsRaRe}
\end{equation}
with $n \approx 3$. The performance of equation~\eqref{eq:NuvsRaRe}, with the limit
Nusselt distributions given in equations~\eqref{eq:NuvsRe} \eqref{eq:NuvsRaRe} is tested
in figure~\ref{fig:NuCf2d}(a) against a comparison with the DNS data in the mixed convection regime.
The agreement seems to be fair in the whole 
parameter space covered by the simulations, although the formula cannot obviously capture the 
previously noticed slight Nusselt with $\Rey_b$ which
is observed mainly around unity bulk Richardson number.

\begin{figure}
 \centering
  \includegraphics[width=12.cm,clip]{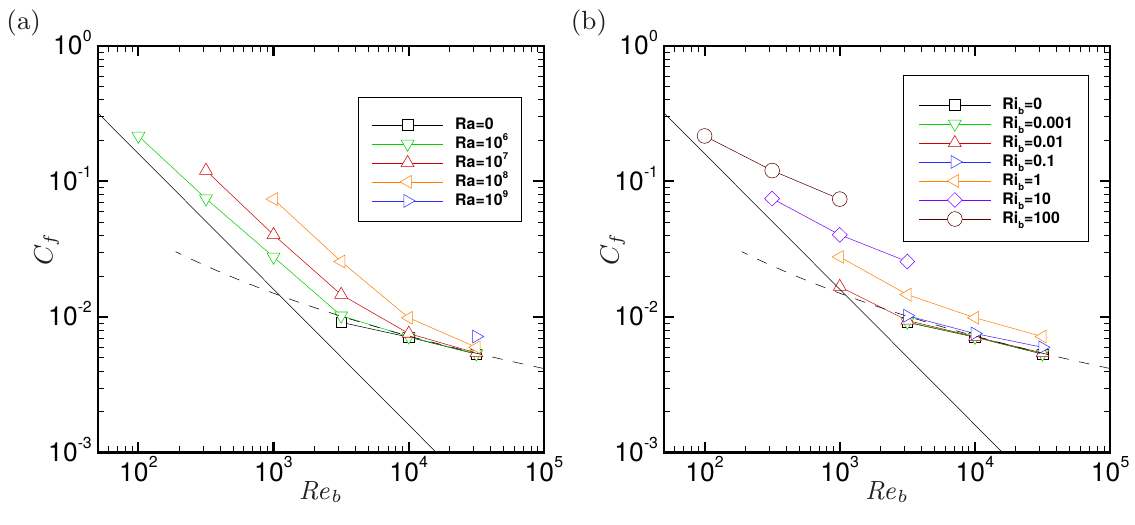}
  \caption{Distributions of friction coefficient as a function of Reynolds number, grouped into lines with same $\Ra$ (a) and with same $\Ri_b$ (b).
           The solid and dashed lines in (b) indicate the laminar Poiseuille curve $C_f = 16/\Rey_b$, and Prandtl's friction law for turbulent channels (equation~\eqref{eq:CfPrandtl}), respectively.}
 \label{fig:Cf}
\end{figure}

The distribution of the friction coefficients, defined as
\begin{equation}
C_f = 2 \tau_w / (\rho u_b^2), \label{eq:Cf}
\end{equation}
is shown in figure~\ref{fig:Cf}, grouped into curves with constant $\Ra$ (a) and with constant $\Ri_b$ (b).
For the sake of reference, in the figure we also show the friction curve for laminar channel flow, and
Prandtl's turbulent friction law, derived by solving the equation
\begin{equation}
\sqcf = \frac 1k \log \left( \frac {\Rey_{b}}2 \sqcfi \right) + C - \frac 1k, \label{eq:CfPrandtl}
\end{equation}
with $k=0.383$, $C=4.17$~\citep{pirozzoli_14}. We find (panel (a)) that increasing the Rayleigh number 
consistently yields an increase of $C_f$, which however tends to saturate at high $\Rey_b$.
In the low-$\Rey_b$ limit the friction curves appear to be nearly parallel to the 
laminar curve, although the $\Ra$-dependent displacement suggest that the structure
of the velocity field is different from the classical Poiseuille representation, as we previously pointed out.
The same data reported at constant $\Ri_b$ in panel (b) show a steady upward displacement of the
friction curves with $\Ri_b$, which is suggestive of a possible parametrization of the 
friction coefficient in the form
\begin{equation}
C_f(\Ra,\Rey_b) = C_f(\Rey_b)+\Delta C_f(\Ri_b), \label{eq:Cffit}
\end{equation}
with ${C_f}(\Rey_b)$ given in equation~\eqref{eq:CfPrandtl}.
Fitting the DNS data we obtain the following empirical representation for the
correction due to buoyancy
\begin{equation}
\Delta C_f(\Ri_b) = \left( 1+ 2 \Ri_b \right)^{0.3}. \label{eq:DeltaCf}
\end{equation}
The performance of the predictive formulas \eqref{eq:Cffit}+\eqref{eq:DeltaCf} 
can be appreciated in figure~\ref{fig:NuCf2d}(b), where $C_f$ is shown in the $\Ra$-$\Rey_b$ plane.
It appears that this simple parametrization correctly captures the increasing trend of $C_f$ 
in the presence of finite buoyancy, although quantitative agreement with DNS data is not perfect
under all flow regimes.

%It is noteworthy that previous investigators~\citep{iida_97,sid_15} quote the presence
%of a minimum of the friction coefficient in numerical simulations  

\section{Conclusions} \label{sec:concl}

We have carried out direct numerical simulations of turbulent channel flows with 
unstable thermal stratification in a wide range of Reynolds and Rayleigh numbersvarying  between the extreme cases of pure free and 
forced convection. Concerning the large--scale structures, the most interesting effect of mixed
convection is the formation of quasi--longitudinal rollers which fill the entire channel height, 
and whose spanwise aspect--ratio may be very large, probably depending to some extent on the size of the
computational box. It is worthwhile noting that this effect, absent
in pure Rayleigh--B\'enard convection and in the turbulent channel flow, shows--up for a wide range of Richardson numbers
(based on DNS data, we find at least $0.01 \le \Ri_b \le 100$), hence virtually in all
situations where mixed convection is relevant. The core rollers can be interpreted as the
turbulent counterpart of Rayleigh instability modes, which become ordered under the action of the mean shear.
It should be noted, however, that the laminar rollers have a typical aspect--ratio of unity,
whereas turbulent rollers are typically much more oblate. Another interesting feature of 
turbulent rollers is their tendency to meander in the spanwise direction, again in
a fashion reminiscent of the wavy secondary instabilities of laminar rollers.
Maximum meandering seems to occur at $\Ri_b \approx 1$, whereas maximum ordering
is found in conditions close to pure forced convection, namely $\Ri_b \approx 0.01$.
The near--wall flow organization consists of the typical pattern of thermal plumes 
at $\Ri_b \gtrsim 1$, and of momentum streaks at $\Ri_b \lesssim 1$, with strong modulating influence 
from the core rollers.

While the flow statistics are well understood in the case of pure forced convection, 
things are nor clear--cut as the limit of free convection is approached. 
Specifically, in--depth analysis of the flow statistics shows
that the mean temperature is far from the $y^{-1/3}$ scaling predicted by Prandtl
based on the assumption that the flow is dominated by wall--attached plumes.
Reasons for this discrepancy may be related to the importance of the core rollers in 
global redistribution of temperature which destroy strict wall scaling. Based on the present dataset, however, we cannot
rule out the possibility that Prandtl's scaling only emerges at Rayleigh numbers much higher
than those considered in this paper and currently accessible to DNS.
The same arguments are likely to apply to the horizontal velocity fluctuations, whose variance
is decreasing with the wall distance, rather than increasing according to
 $y^{2/3}$ as required by
plume--based scaling. On the other hand, vertical velocity and temperature fluctuations 
closely conform to the model predictions, hence suggesting that vertical motions are 
mainly controlled by thermal plumes.
These findings have large impact on classical parametrizations of heat flux and wall
friction, mainly based on the Monin--Obukhov dimensional ansatz that wall--scaled mean flow gradients
and fluctuations should be universal functions of $y/L$. Based on the present DNS database,
we find that this assumption is satisfied with reasonable accuracy,
probably acceptable for `first--order' estimates. However, we also find that the predictive 
accuracy of empirical formulas relying on Monin--Obukhov universality varies 
considerably depending on the variables. Specifically, we find that the individual profiles of 
vertical velocity, temperature variance and the $u$--$\theta$ correlation
follow quite well the universal trends. On the other hand, the profiles of mean temperature and velocity 
as well as the horizontal velocity variances only follow the theoretical 
trends in a global sense, whereas individual profiles show a radically different behaviour.
This is likely to result from the coexistence of a plume--dominated scaling with statistics
varying with $y/L$, and a core--dominated scaling, with quantities scaling as $h/L$~\citet{panofsky_77}. 
Direct verification of this inference would require simulations covering individually a wide 
range of $y/L$, which is only possible using simultaneously extreme values of $\Ra$, $\Rey_b$,
and certainly beyond the current DNS capabilities.
An relevant outcome of the present study is a set of modified Businger--Dyer relationships 
for the various flow variables based on fitting the DNS data.
Differences with curve fits of current use in meteorological parametrizations are 
generally small, except for the mean streamwise velocity, 
for which we recover a very mild $(y/L)^{-1/7}$ variation in the light wind regime,
which is sensibly different from the $(y/L)^{-1/4}$ of current use.
This finding is potentially interesting 
as correct parametrization of the light--wind regime
can have large impact on the prediction of weather conditions featuring strong thermal
instability, as is the case of the Indian monsoon circulation \citep{rao_06}.
The present DNS data are especially important for this purpose, 
as mean wind data in the light wind
regime are extremely scattered owing to difficulties inherent
to field atmospheric measurements.
We should however also recall that the high--$\Ri_b$ regime
is particularly challenging also for DNS, as it corresponds to the 
bottom--right corner of figure~\ref{fig:mesh}, where Rayleigh numbers
are high but Reynolds number are rather low, thus possibly casting uncertainties
on direct applicability of DNS data to the context of atmospheric turbulence.

Of large practical interest is also the prediction of friction and 
heat transfer as a function of the bulk flow parameters.
Regarding heat transfer, a peculiar behavior is observed whereby the 
addition of bulk mass flow initially leads to a decrease of the Nusselt number
down to $\Ri_b \approx 1$, and then an increase moving toward the 
forced convection regime.
Reasonable estimates for $\Nu$, which however do not incorporate this effect are
obtained by simple geometric averages of the values found
in the limits of free and forced convection.
Empirical corrections to the classical heat transfer formulas valid for neutral channels
are also established to incorporate the effect of buoyancy by fitting the DNS data.
A simple parametrization based on the sole bulk Richardson number
yields predictions with $O(10\%)$ accuracy in the whole $\Ra, \Rey_b$ parameter space.
 
%Flow statistics are available at the web page
%{\tt http://newton.dma.uniroma1.it/mixed\_conv/},
%with supporting documentation.

\begin{acknowledgments}
This work was carried out on the national e-infrastructure of
SURFsara, a subsidiary of SURF cooperation, the collaborative ICT
organization for Dutch education and research. We also acknowledge
PRACE for awarding us access to FERMI based in Italy at CINECA under
PRACE project numbers 2015133124 and 2015133204. 
We also acknowledge that the results
of this research have been achieved using the DECI resource Archer
based in the United Kingdom at Edinburgh with support from the PRACE
aisbl.
\end{acknowledgments}
\bibliographystyle{jfm}
\bibliography{references}
\end{document}